\documentclass[final,3p,times,authoryear]{elsarticle}

\usepackage{amssymb}

\usepackage{graphics} 
\usepackage{graphicx}
\usepackage{array}
\usepackage{amsmath} 
\usepackage{amssymb}  
\usepackage[caption=false]{subfig}
\usepackage{algorithmic}
\usepackage[plain]{algorithm}
\usepackage{mathrsfs} 
\usepackage{appendix}
\usepackage{bm,upgreek}
\usepackage[normalem]{ulem} 
\usepackage{caption}

\usepackage{color}
\usepackage[colorinlistoftodos]{todonotes}

\usepackage[left,modulo]{lineno}

\graphicspath{%
	{/}
	}

\def\inclps#1#2#3{\resizebox{#1}{#2}{\includegraphics{#3}}}

\definecolor{fgred}{rgb}{0.8,0,0}     
\definecolor{fgblue}{rgb}{0,0,1}      
\definecolor{fggreen}{rgb}{0,0.5,0}     

\journal{J. Mech. Phys. Solids}

\begin{document}

\begin{frontmatter}

\title{Finite deformations govern the anisotropic shear-induced area reduction of soft elastic contacts}

\author[label0,label1]{J. Lengiewicz}
\author[label2]{M. de Souza}
\author[label2,label3]{M. A. Lahmar}
\author[label3]{C. Courbon}
\author[label2]{D. Dalmas}
\author[label1]{S. Stupkiewicz\corref{cor1}}
\author[label2]{J. Scheibert\corref{cor2}}
\cortext[cor1]{Corresponding author, sstupkie@ippt.pan.pl}
\cortext[cor2]{Corresponding author, julien.scheibert@ec-lyon.fr}
\address[label0]{{Department of Engineering, Faculty of Science, Technology and Medicine, University of Luxembourg, Luxembourg}}
\address[label1]{Institute of Fundamental Technological Research (IPPT), Polish Academy of Sciences, Pawi\'nskiego 5B, 02-106, Warsaw, Poland}
\address[label2]{Univ Lyon, Ecole Centrale de Lyon, ENISE, ENTPE, CNRS, Laboratoire de Tribologie et Dynamique des Syst\`emes LTDS, UMR 5513, F-69134 Ecully, France}
\address[label3]{Univ Lyon, Ecole Centrale de Lyon, ENISE, ENTPE, CNRS, Laboratoire de Tribologie et Dynamique des Syst\`emes LTDS, UMR 5513, F-42100, Saint-Etienne, France}

\begin{abstract}
Solid contacts involving soft materials are important in mechanical engineering or biomechanics. Experimentally, such contacts have been shown to shrink significantly under shear, an effect which is usually explained using adhesion models. Here we show that quantitative agreement with recent {high-load} experiments can be obtained, with no adjustable parameter, using a non-adhesive model, provided that finite deformations are taken into account. Analysis of the model uncovers the basic mechanisms underlying shear-induced area reduction, local contact lifting being the dominant one. We confirm experimentally the relevance of all those mechanisms, by tracking the shear-induced evolution of tracers inserted close to the surface of a smooth elastomer sphere in contact with a smooth glass plate. Our results suggest that finite deformations are an alternative to adhesion, when interpreting a variety of sheared contact experiments involving soft materials.
\end{abstract}




\end{frontmatter}



\section{Introduction}
\label{intro}

Rough contacts are ubiquitous in both natural and engineering systems, and indeed, rough contact mechanics have been actively investigated in the last decades (see e.g.~\citet{vakis_modeling_2018} for a recent review). Most of the effort has been devoted to the normal contact of frictionless interfaces (see~\citet{muser_meeting_2017} for a comparison of various modelling approaches to such a problem). However, recent experiments involving soft materials like polymers or human skin have revealed complex changes to the contact morphology when a frictional rough contact is submitted to an additional shear load. Not only is the overall real contact area significantly reduced \citep{sahli_evolution_2018, weber_frictional_2019}, but it also becomes increasingly anisotropic \citep{sahli_shear-induced_2019}, two effects that have not been satisfactorily explained yet. For both effects, smooth sphere/plane contacts have been shown to obey similar behaviour laws as rough contacts \citep{sahli_evolution_2018,sahli_shear-induced_2019}. It is thus appealing to start investigating anisotropic shear-induced area reduction in such simpler, single sphere/plane contacts between smooth solids.

Experimentally, elastomeric sphere/plane contacts are known to evolve from a circular area under pure normal load to a smaller, ellipse-like area in macroscopic sliding regime \citep{savkoor_effect_1977, waters_mode-mixity-dependent_2010, petitet_materiaux_2008, sahli_evolution_2018, sahli_shear-induced_2019, mergel_continuum_2019}. So far in the literature, such an area reduction during incipient tangential loading  of a sphere/plane contact has been interpreted using adhesion models, most of which are based on linear elastic fracture mechanics~\citep{savkoor_effect_1977, johnson_continuum_1996, johnson_adhesion_1997, waters_mode-mixity-dependent_2010, ciavarella_fracture_2018, papangelo_mixed-mode_2019, papangelo_shear-induced_2019}. Those fracture-based models start with a JKR (Johnson-Kendall-Roberts, \citet{johnson_surface_1971}) description of a frictionless, adhesive, linear elastic spherical contact, for which the pressure field at the contact's periphery is locally that of a mode I (opening) {crack}. The models continue noting that applying a tangential force $Q$ to a fully stuck contact introduces a shear stress field at the contact's periphery which is that of a mode II (shear) {crack}. The contact radius $a$ is then obtained by equating the available mechanical energy at the contact's {periphery}
to a relevant fracture energy. In the oldest such model \citep{savkoor_effect_1977}, the fracture energy was simply taken as {the work of adhesion of the interface,} $w_0$. This led to an area reduction much larger than observed experimentally, suggesting that the effective fracture energy is much larger than $w_0$. Subsequent models thus considered a fracture energy given by {$w_0 f$}, with $f>$1 a mode-mixity function describing the interaction between adhesion and friction at the crack tip, and accounting for the increased energy dissipation due to, for instance, micro-slip within the sheared contact.

Fine-tuning of the amplitude and shape of {$f$} allowed to quantitatively reproduce experimental results on the area reduction of smooth PDMS (PolyDiMethylSiloxane) spheres in contact against smooth glass plates \citep{ciavarella_fracture_2018, papangelo_mixed-mode_2019}. Accounting for a second mode-mixity function describing the interaction between mode I and mode III (antiplane shear) further allowed to reproduce the anisotropic properties of the area changes \citep{papangelo_shear-induced_2019}. However, such agreement required at best to fit the amplitude of {$f$}~\citep{ciavarella_fracture_2018, papangelo_mixed-mode_2019}, at worst to interpolate the whole {$f$} functions from one experiment out of the set to be reproduced~\citep{papangelo_shear-induced_2019}. This limitation arises because our current understanding of the physical mechanisms lumped into the mode-mixity functions, and more generally of the interactions between adhesion and friction, remains unsatisfactory. {For} the same reason, other adhesion models, like the numerical ones of~\citet{mergel_continuum_2019, kajeh-salehani_modeling_2019,mergel_contact_2020} or the theoretical one of~\citet{mcmeeking_interaction_2020}, also require ad hoc descriptions of the local interfacial behaviour under coupled normal and tangential loading.

Surprisingly, simpler, adhesionless models based only on elasticity have not been proposed. This is partly historical, because the first experiments on shear-induced contact reduction \citep{savkoor_effect_1977, waters_mode-mixity-dependent_2010} were mainly performed under small (or even negative) normal loads, for which adhesive stresses are expected to dominate those due to indentation. Recent experiments at much higher normal loads \citep{sahli_evolution_2018}, for which adhesive stresses are expected to be less prominent, suggest that the area reduction may also occur in the absence of adhesion. Another possible reason for having overlooked elastic models is the nature of the materials used for the experiments. They mainly considered elastomers, which are nearly incompressible, in contact with a rigid substrate. In those conditions, the normal and shear stresses at the contact interface between linear elastic solids are uncoupled~\citep{johnson_contact_1985}, so that the observed effect of the tangential load on the contact area is unexpected.

Here, we investigate the hypothesis that the shear-induced contact area reduction is an elastic effect enabled by the nonlinear, finite-deformation behaviour of the elastomer. Recently, this possibility was suggested in~\citet{wang_sphere_2020}, and some preliminary support was brought by~\citet{mergel_contact_2020} using 2D simulations, but it has never been tested on 3D sphere/plane contacts. To fully test the hypothesis, we have developed a computational 3D model that combines {a hyperelastic bulk and a non-adhesive but frictional interface.}
{The} model suitably fits the macroscopic experimental data from \citet{sahli_evolution_2018,sahli_shear-induced_2019} and quantitatively reproduces the anisotropic evolution of contact area, with no adjustable parameter (Section~\ref{sec:model}). {Our} results strongly suggest that the contact area reduction is governed by the finite-deformation mechanics, with a dominating effect of local contact lifting and less pronounced effect of micro-slip-induced in-plane deformation. {We} then present original experiments in which those two mechanisms can indeed be directly observed and quantified (Section~\ref{sec:experiment}).

\section{Finite-strain modelling of shear-induced contact area reduction}\label{sec:model}

The finite-strain framework and the Tresca friction model are the two essential features of our model, and these are combined with the finite-element method as a suitable spatial discretization scheme. 
Since the individual ingredients of the model are rather standard, perhaps except for some details of the computational treatment, the model is only briefly summarizeded in Section~\ref{sec:computmodel}, and its detailed description is provided in \ref{app:model}.

\subsection{Computational model: finite-strain framework and Tresca friction}\label{sec:computmodel}

\begin{figure}[ht!]
    \centering
    \subfloat[]{\inclps{!}{0.216\textwidth}{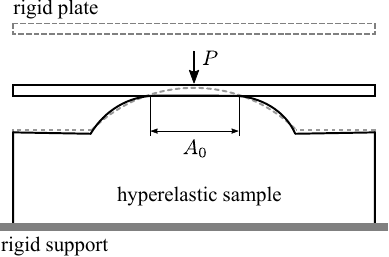}} \hspace*{3em}
    \subfloat[]{\raisebox{2.24ex}{\inclps{!}{0.192\textwidth}{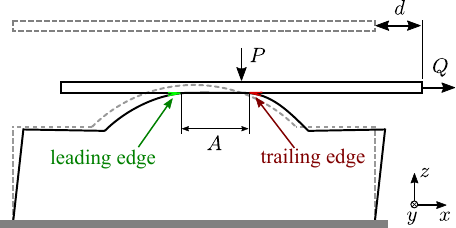}}}
    \caption{2D sketch of the (3D) soft-sphere/rigid-plane contact under study. The initial {(dashed lines) and deformed (solid lines) configurations are shown} for two stages: (a) {under pure normal load $P$} and (b) when an additional  tangential displacement $d$ is applied, giving rise to a tangential load $Q$. {In all subsequent figures, the leading (resp.\ trailing) edge is always on the left (resp.\ right) side.}}
    \label{fig: fem schematics}
\end{figure}

The contact problem under consideration, sketched in Fig.~\ref{fig: fem schematics}, corresponds to the normal and tangential loading of a hyperelastic spherical solid in (adhesionless) frictional unilateral contact with a rigid plate. The finite-strain framework employs the geometrically exact kinematics of finite deformations (\ref{Sec: finite strain}) and contact (\ref{sec:kinematics}), as well as adequate constitutive descriptions of both the bulk elasticity of the sphere (\ref{Sec: finite strain}) and the frictional behaviour at the contact interface (\ref{sec:Tresca}).

Hyperelasticity is treated using the Mooney--Rivlin model, in the nearly incompressible version. Disregarding compressibility and the related bulk modulus, the model (Eq.~\eqref{Eq: MR model}) involves two parameters, $\mu_1$ and $\mu_2$, such that $\mu=\mu_1+\mu_2$, with $\mu$ the shear modulus. In the following, we will consider two particular cases reducing to a single parameter, $\mu$: (i) the neo-Hookean model, for which $\mu_2=0$ and $\mu_1=\mu$, and (ii) an arbitrarily chosen other combination of parameters, namely $\mu_1=\mu_2=\mu/2$ (simply denoted as Mooney--Rivlin from now). Note that $\mu$ can be readily determined for any particular experimental sphere/plane contact from the knowledge of the contact area, $A_0=A(Q=0)$, under a pure, known normal load, $P$.

For a wide range of tribological material pairs, the static friction force of sphere/plane contacts is measured to be proportional to the contact area (see~\citet{sahli_evolution_2018} and references therein). It is thus appealing to examine the simplest friction model that automatically produces such a dependence, namely the Tresca friction model, which is {employed here} (\ref{sec:Tresca}). In the Tresca model, no slip occurs locally until the contact shear strength, $\sigma$, is reached, and $\sigma$ is assumed constant and independent on the normal contact {traction.} The Tresca model, even if oversimplified, is expected to provide a reasonable approximation of the tangential stress distribution (at least) when the static friction force, $Q_{\rm s}$, is reached. Note that $\sigma$ can be measured in all experiments that monitor the contact area \citep[e.g.,][]{savkoor_effect_1977,waters_mode-mixity-dependent_2010,sahli_evolution_2018,mergel_continuum_2019}, as the ratio of $Q_{\rm s}$ over $A_{\rm s}=A(Q=Q_{\rm s})$.

The finite-element treatment of the contact problem at hand is described in {\ref{app:fe-model} and in Section~S.1.1 in Supplementary Information}. Note that the Tresca friction model may lead to significant convergence problems, particularly when the interfacial shear strains are large, which will be the case in our simulations. To circumvent those problems, a Prakash-Clifton-like regularization scheme (\ref{sec:Tresca}) has been employed, which allowed our actual computations to be successfully carried out for a relatively fine finite-element mesh. 

We have applied our model to perform a direct quantitative comparison with the PDMS-sphere/glass-plate experiments reported in~\citet{sahli_evolution_2018, sahli_shear-induced_2019}. In particular, the geometry, boundary conditions and loading conditions match the experimental ones. The elastomer sample (identical to that in inset of Fig.~\ref{fig:Setup_scretch}) is a {cylinder, the top of which features a spherical cap} with a radius of curvature of 9.42\,mm. It is fixed on a rigid support. The spherical cap is first brought into normal contact with a rigid plate under constant normal load $P$ (we chose four representative values of $P$ covering the whole range explored in~\citet{sahli_evolution_2018}: 0.27, 0.55, 1.65 and 2.12\,N). The contact is then sheared by pulling the plate horizontally with a constant velocity $V=0.1$\,mm/s.

A value of $\mu=0.60$\,MPa was derived in a unique manner using the experimental values of $P$ and {$A(Q=0)$ (see} \ref{app:fe-model}). Similarly, a value of $\sigma=0.41$\,MPa was derived using the experimental values of $A_{\rm s}$ and $Q_{\rm s}$. Note that the ratio $\sigma/\mu$ is significant, such that shear strains exceeding 50\% are expected. This is far beyond the range of validity of the small-strain theory, and linear elasticity in particular, so that the finite-strain framework used here is actually essential. {At those large strains, the mechanical behaviour of the elastomer used in the experiments of~\citet{sahli_evolution_2018,sahli_shear-induced_2019} (Sylgard 184) is already well beyond its linear range~\citep{nguyen_surface_2011,maraghechi_experimental_2020}.}

{To further characterize the contact conditions, let us introduce the Hertzian contact radius $a_{\rm H}=(3PR/(16\mu))^{1/3}$, where the Poisson's ratio $\nu=0.5$ has been assumed for simplicity. 
In the present conditions, the dimensionless ratio $a_{\rm H}/R$ varies between 0.098 and 0.195 for $P=0.27$\,N and $P=2.12$\,N, respectively, thus exceeding the usual range of validity of the Hertz contact theory, $a_{\rm H}\ll R$ \citep{johnson_contact_1985}.} 
{For the shear loading, the relevant dimensionless parameter is $\sigma/\mu$.}

\begin{figure}[ht!]
    \centerline{
    \begin{tabular}{cc}
      \raisebox{-10ex}{
      {\footnotesize (a)}~\inclps{0.48\textwidth}{!}{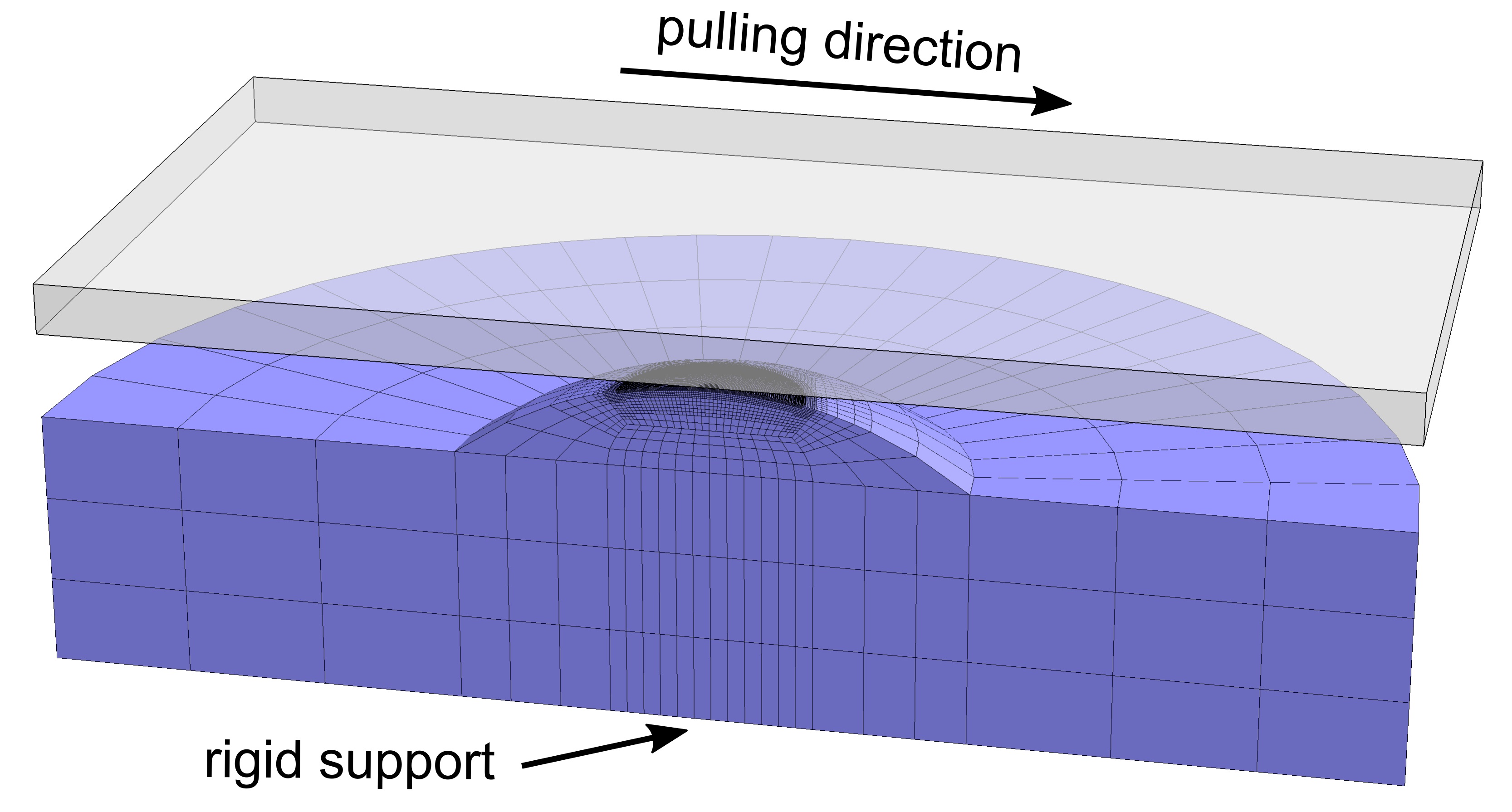}
      } &
      \begin{tabular}{c}
        {\footnotesize (b)}~\inclps{0.43\textwidth}{!}{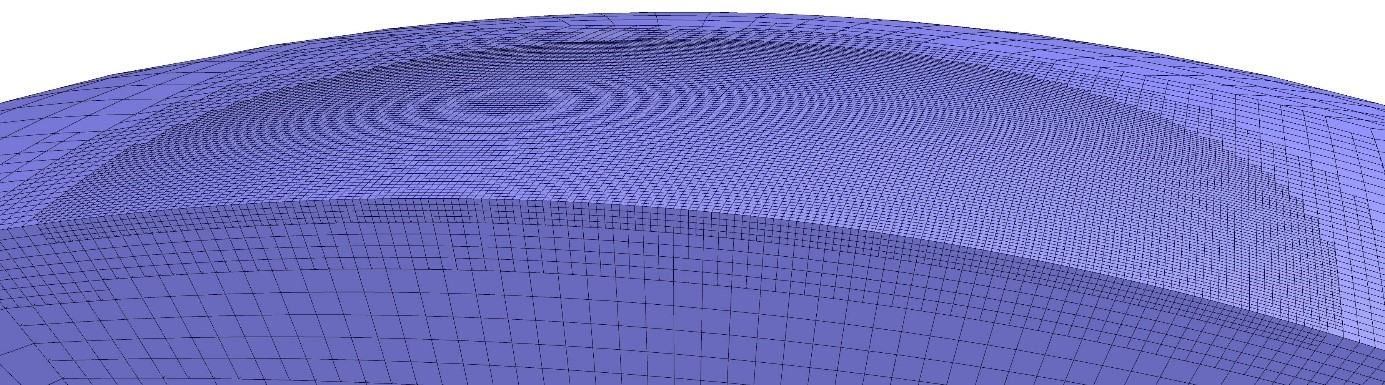} \\
        {\footnotesize (c)}~\inclps{0.43\textwidth}{!}{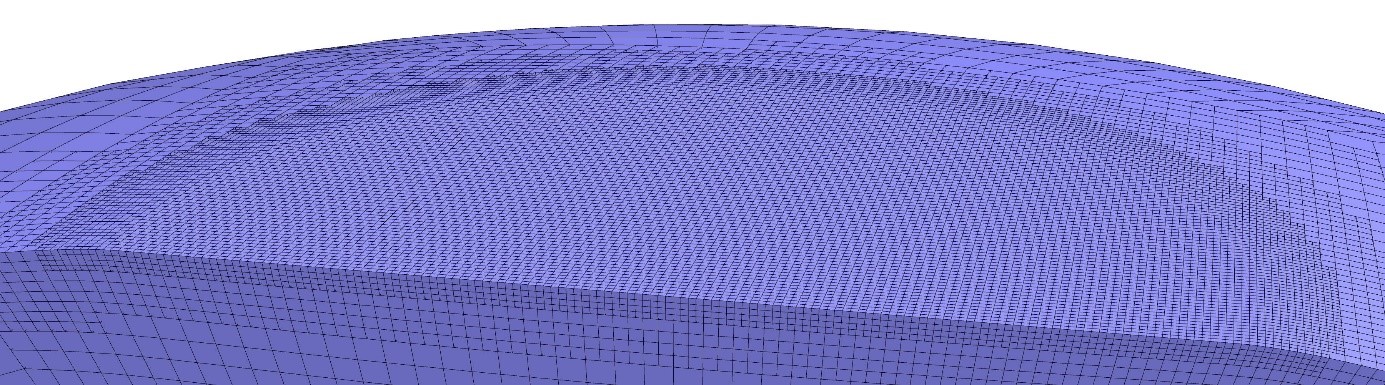} \\
        {\footnotesize (d)}~\inclps{0.43\textwidth}{!}{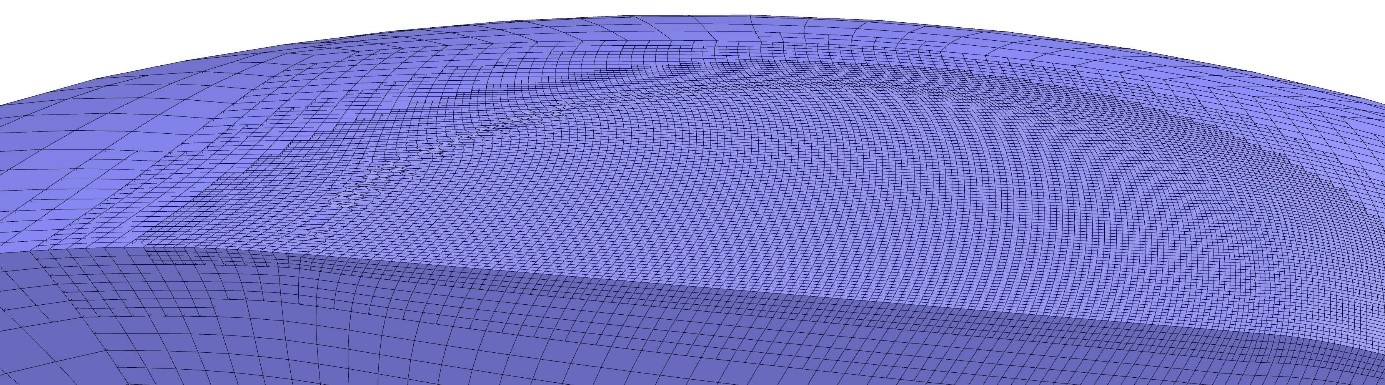}
      \end{tabular}
    \end{tabular}
    }
    \caption{Finite-element mesh used in the case of the highest normal load $P=2.12$\,N: (a) undeformed mesh, (b) zoom-in of the undeformed mesh, (c) zoom-in of the deformed mesh after initial normal loading, (d) zoom-in of the deformed mesh at full sliding. {The leading edge is on the left.}}
    \label{fig:3d fem mesh}
\end{figure}

Figure~\ref{fig:3d fem mesh} shows the finite-element mesh used in the case of $P=2.12$\,N (note that the size of the refined-mesh region is adjusted to the normal load), and zoom-ins for select configurations. In particular, panel (d) illustrates the deformation pattern for $Q=Q_{\rm s}$, showing a significant shear deformation in the subsurface layer and a non-uniform in-plane deformation of the contact surface. The corresponding high mesh distortion, visible at the leading {edge, is} due to the jump of the tangential contact traction, introduced by the Tresca model, and to the non-conforming discretization of the contact area boundary (see related discussion in {Section~S.1.2}).

\subsection{Shear-induced contact area reduction}\label{sec:reductionresults}

\begin{figure}[ht!]
  \centerline{
  \begin{tabular}{cc}
  \inclps{0.50\textwidth}{!}{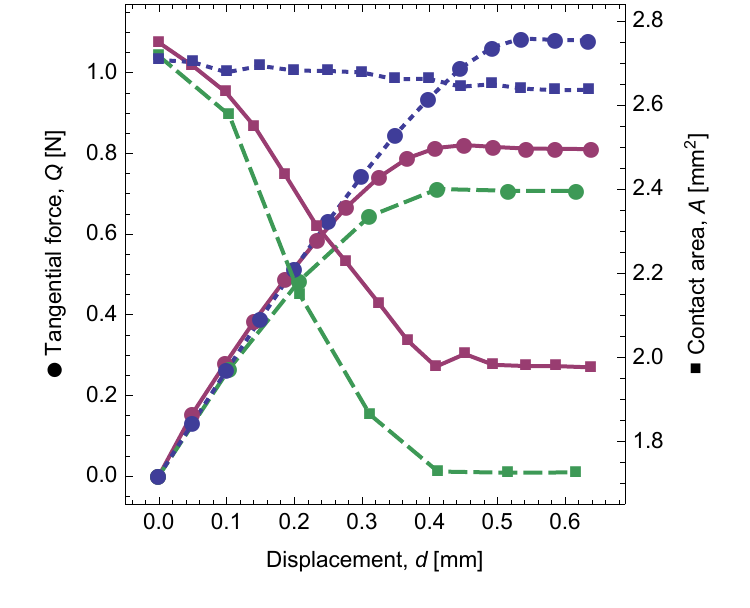} &
  \raisebox{1.3ex}{\inclps{0.39\textwidth}{!}{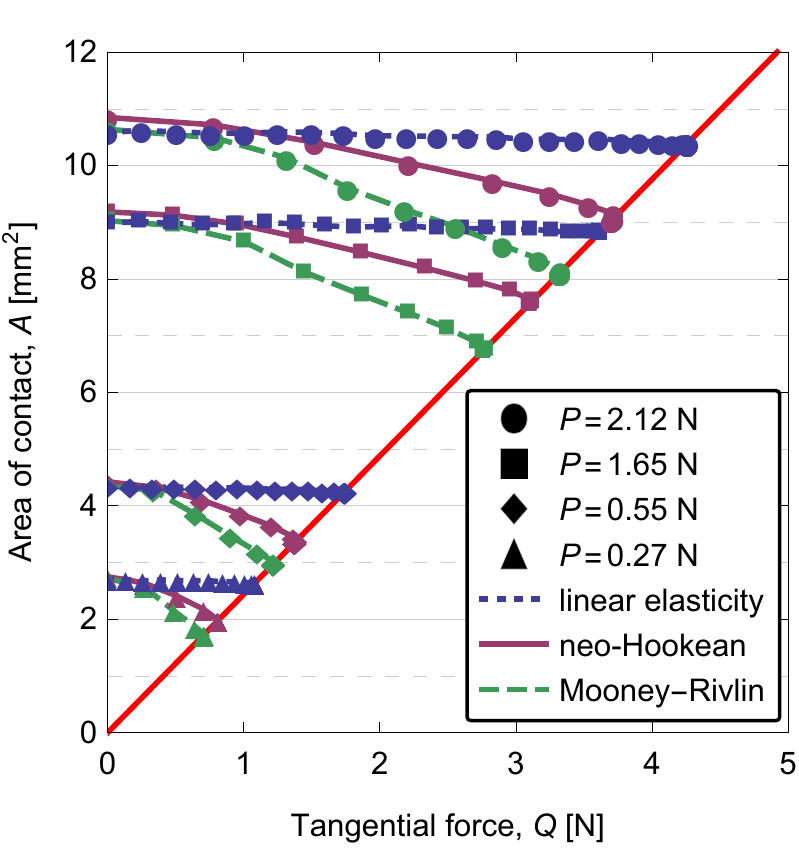}} \\[-1ex]
  {\footnotesize (a)} & ~~~~{\footnotesize (b)}
  \end{tabular}
  }
    \caption{Shear-induced contact area reduction predicted for various elastic models. (a) Concurrent evolution of the tangential force, $Q$ (dots), and the contact area, $A$ (squares), as a function of the rigid plate displacement, $d$, for three different elastic models: linear elastic (dotted blue), neo-Hookean (solid purple) and Mooney--Rivlin (dashed green). $P=0.27$\,N. (b) $A$ vs $Q$, for the three elastic models and for the four normal loads. The markers correspond to selected instants; the computations were carried out with a much finer time stepping. {Solid straight red line: $Q=\sigma A$.}}
    \label{fig:area reduction}
\end{figure}

Using the computational model described above, we simulated the shear-induced contact area reduction in the sphere/plane experiments of~\citet{sahli_evolution_2018,sahli_shear-induced_2019}, for the two hyperelastic models (neo-Hookean and Mooney--Rivlin). Typical results are reported in Fig.~\ref{fig:area reduction}(a) which shows the concurrent evolutions of the tangential force, $Q$, and the contact area, $A$, as a function of the displacement of the rigid plate, $d$. For comparison, the results corresponding to linear elasticity are also included in Fig.~\ref{fig:area reduction}, even if the range of strains expected in our conditions does not really admit application of the small-strain framework.

Figure~\ref{fig:area reduction}(a) shows that the three models yield drastically different predictions, both for the tangential force and for the contact area. While the initial contact area at zero tangential force is hardly affected by the model, major differences appear as the tangential force increases. A significant area reduction is predicted for both hyperelastic models, but it remains negligible for the linear elastic model. This result already indicates that the finite-deformation framework is actually a prerequisite to reproduce an area reduction of several tens of percents, as observed experimentally. Note that the amplitude of the reduction is larger for the Mooney--Rivlin model (37\% for $P=0.27$\,N) than for the neo-Hookean model (28\%), showing that the effect of the material model that governs the hyperelastic behaviour of the sample is significant. Those very different final contact areas, combined with the very same contact shear strength, $\sigma$, explain why the final friction force is significantly larger for the linear elastic model, and is also different for both hyperelastic models.

Figure~\ref{fig:area reduction}(b) synthesizes all results by showing, for all normal loads and all three elastic models, the contact area $A$ as a function of the tangential force $Q$. The absence of area reduction for the linear elastic model and the difference in its amplitude for the two hyperelastic models is clearly evidenced. The fact that all curves end on the red line indicates that a homogeneous shear stress distribution equal to the contact shear strength $\sigma$ is suitably enforced at the onset of gross {sliding.}

\begin{figure}[ht!]
    \centerline{
    \hspace*{2em}
    \begin{tabular}{cccc}
    \hspace{-1.2em}\raisebox{1.3em}{\inclps{!}{0.22\textwidth}{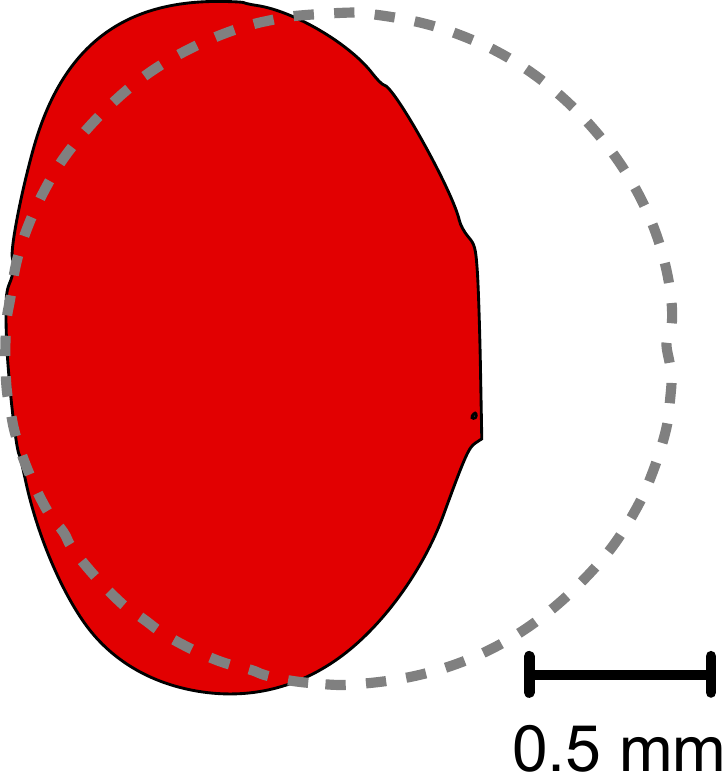}} &
    \hspace{0em}\raisebox{1em}{\inclps{!}{0.24\textwidth}{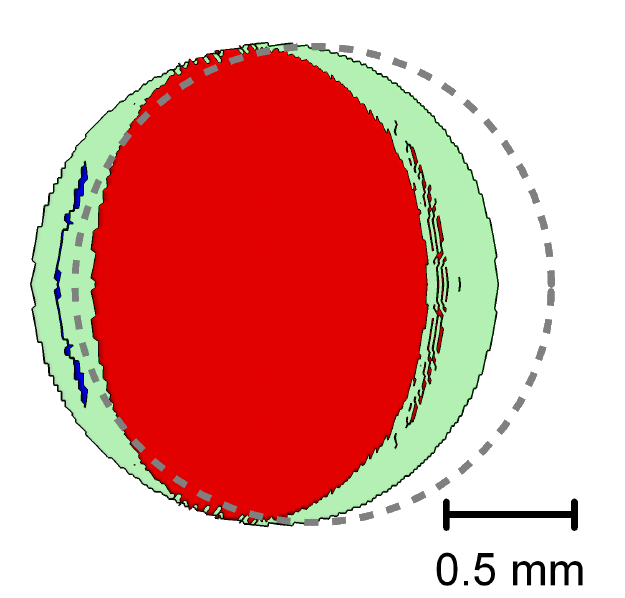}} &
    \hspace{-0.5em}\raisebox{1em}{\inclps{!}{0.24\textwidth}{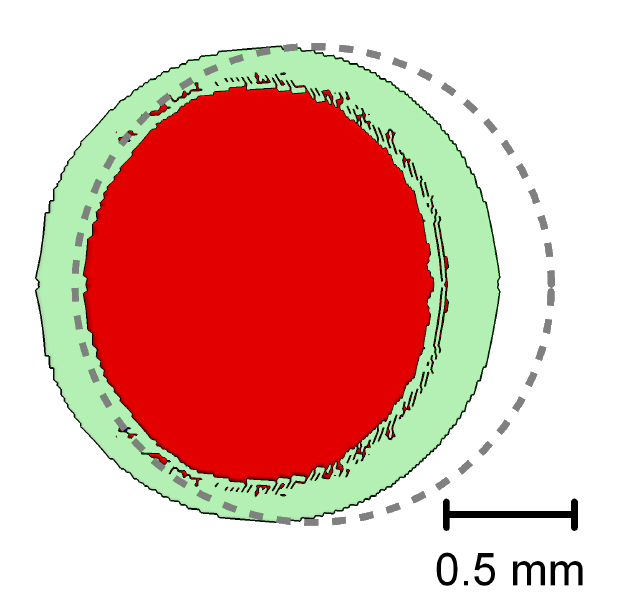}} &
    \hspace{-0.5em}\raisebox{1em}{\inclps{!}{0.24\textwidth}{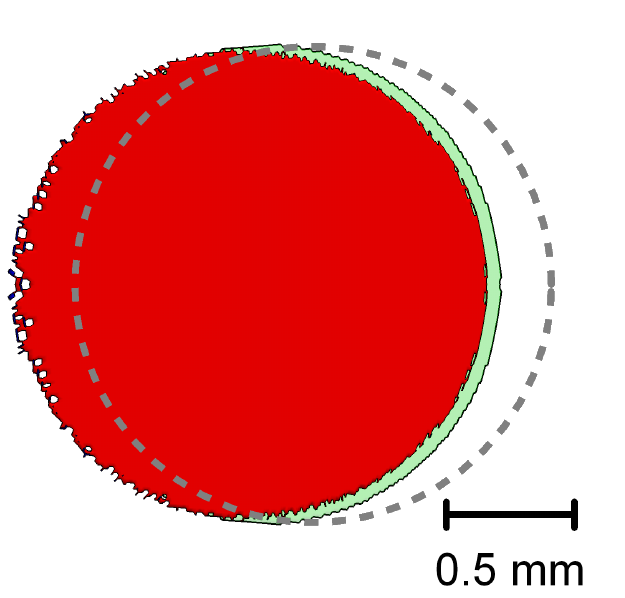}} \\[-2.5ex]
    {\footnotesize (a)} & {\footnotesize (b)} & {\footnotesize (c)} & {\footnotesize (d)}
    \end{tabular}
    }
    \caption{Contact zone (shown in red in the frame attached to the moving rigid plate) at the onset of full sliding for $P=0.27$\,N. (a) Experiment~\citep{sahli_evolution_2018} (this specific diagram has not been published in the reference). (b) Neo-Hookean model. (c) Mooney--Rivlin model. (d) Small-strain linear elastic model. Dashed circles indicate the boundary of the initial contact zone, green regions denote the lifted area in the current (deformed) configuration. {The leading edge is on the left.} {The mesh-dependent features and secondary contact regions visible in panels (b)--(d) are commented in \ref{app:validation} and in Section~S.1.2.}}
    \label{fig:contact zone different models 0.27}
\end{figure}

The differences between the various elastic models concern also the shape of the contact zone. In Fig.~\ref{fig:contact zone different models 0.27}(b)--(d), typical model predictions for the final contact shape are shown for the three elastic models (red regions). Although all models start with an almost identical circular contact when $Q=0$ (dashed lines), the final contact shape has significant differences. The linear elastic model predicts a final contact which remains circular, with almost the same area as in the initial configuration, consistently with the results of Fig.~\ref{fig:area reduction}. The large area reduction in the Mooney--Rivlin model occurs while keeping an essentially circular contact shape. In contrast, the neo-Hookean model leads to an ellipse-like contact shape, with the size reduction occurring almost exclusively along the shear {direction.}

\begin{figure}[ht!]
    \centerline{
    \begin{tabular}{ccc}
	\inclps{0.40\textwidth}{!}{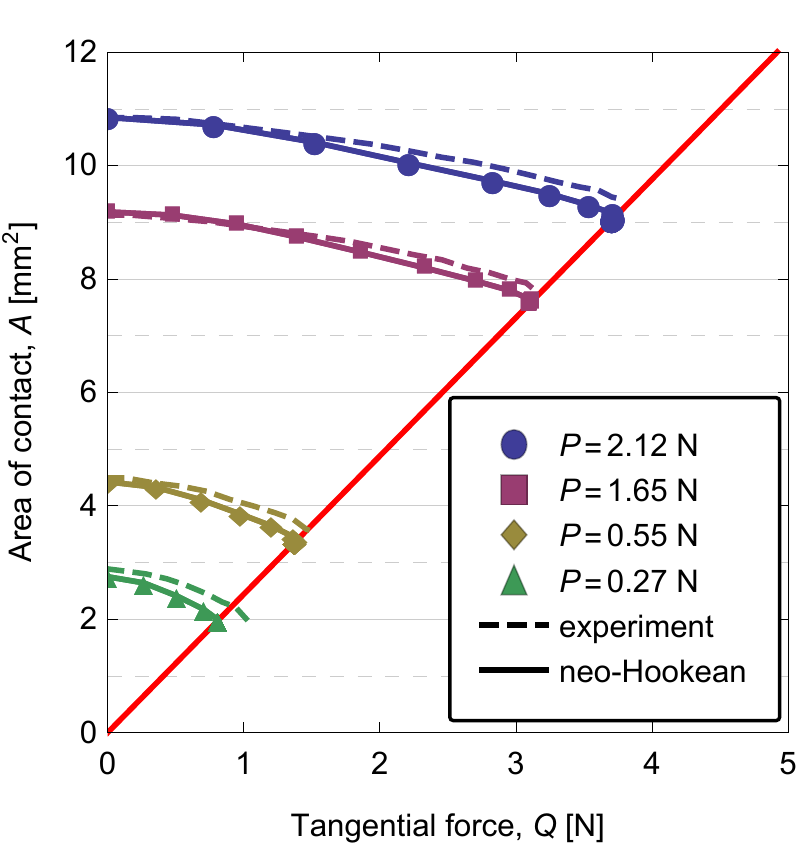} & &
	\inclps{0.40\textwidth}{!}{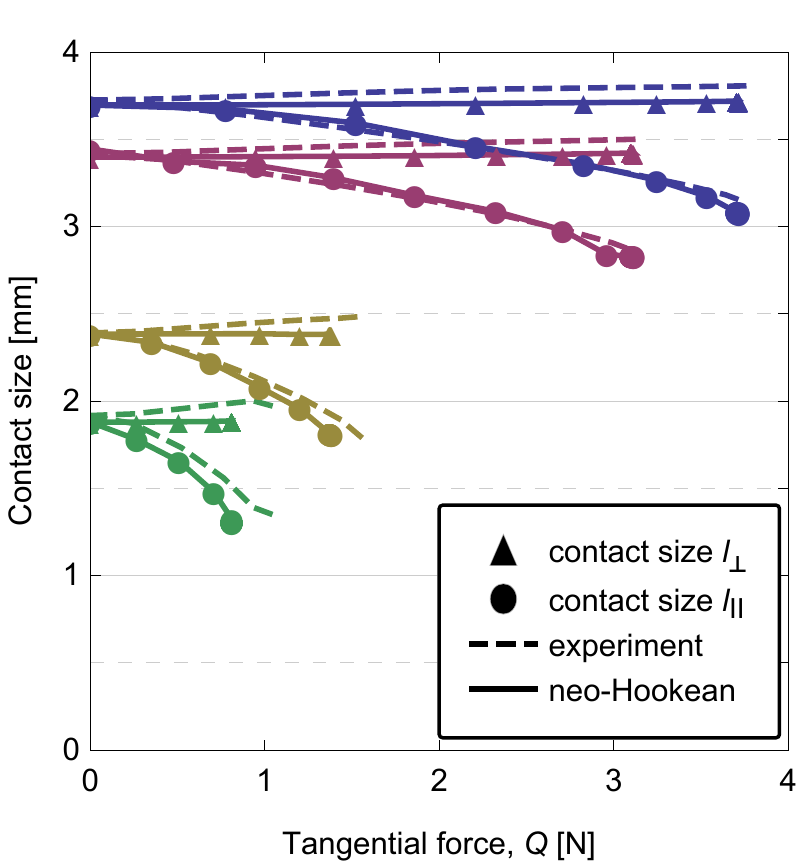} \\
	\hspace*{2em}{\footnotesize (a)} & &
	\hspace*{2em}{\footnotesize (b)}
    \end{tabular}
    }
    \caption{Comparison of the model predictions obtained for the neo-Hookean model, for all normal loads, to the experimental results of \citet{sahli_evolution_2018,sahli_shear-induced_2019}. (a) {$A$ vs $Q$}. {Solid straight red line: $Q=\sigma A$}. (b) Contact size along ($\ell_{\parallel}$) and perpendicular to ($\ell_{\bot}$) the shear loading direction {vs} $Q$.}
    \label{fig:comparison}
\end{figure}

We are now in a position to compare quantitatively our model results to the experimental results of~\citet{sahli_evolution_2018,sahli_shear-induced_2019}. In Fig.~\ref{fig:contact zone different models 0.27}, panel (a) shows the experimental counterpart of panels (b)--(d). It clearly appears that the model that most closely matches the final shape in the experiments is the neo-Hookean model. In particular, both the size and eccentricity of the ellipse-like contact shape is very-well captured. {In the experiment, the radius of curvature of the trailing edge is visibly smaller than that of the leading edge. In the model, this effect is also visible, but less pronounced. Also note that, in the experiments, the final contact region extends beyond the initial contact at the leading edge, an effect which is weaker in the neo-Hookean model, and that will be further commented in Section~\ref{sec:experiment}.}

The excellent agreement of the neo-Hookean model with experiments is further demonstrated in Fig.~\ref{fig:comparison}. Figure~\ref{fig:comparison}(a) directly compares the predicted $A(Q)$ curves with those of~\citet{sahli_evolution_2018} (see their Fig.~2C), while Fig.~\ref{fig:comparison}(b) compares the predicted contact sizes along and orthogonal to shear, $\ell_{\parallel}$ and $\ell_{\perp}$ respectively, with those of~\citet{sahli_shear-induced_2019} (see their Fig.~3b). In both cases, the amplitudes and shapes of the curves are well captured, although the model slightly underestimates $A$ and does not capture the slight increase of $\ell_{\perp}$ observed in the experiments. Note that a similarly good agreement has been obtained for an alternative model with different regularizations and numerical implementations (see {\ref{app:validation} and Section~S.1.3}), thus showing the robustness of our results.

\subsection{Elementary mechanisms of contact area reduction}
\label{sec:mechanisms}

In Section~\ref{sec:reductionresults}, we have shown that our neo-Hookean model provides, without any adjustable parameter, a very good quantitative prediction of the shear-induced contact area reduction observed in experiments. On this model only, we will now perform a thorough analysis of the simulation results to understand what are the elementary mechanisms responsible for such a reduction.

Important qualitative insights can already be obtained from a careful inspection of Fig.~\ref{fig:contact zone different models 0.27}(b). First, the green {region} around the final contact zone {corresponds} to all nodes that were initially in contact, and which are not anymore at the onset of macroscopic sliding. Those nodes have been lifted out of contact during the incipient loading phase, due to the shear-induced deformations of the elastic sphere. {This first elementary mechanism of area variations will be named below} ``contact lifting'' (or simply ``lifting''). Second, small parts (hardly visible in Fig.~\ref{fig:contact zone different models 0.27}(b)) of the final contact are found outside the {deformed} initial contact region (shown as a green region in the current configuration). {These nodes} were initially out of contact, but came into contact during the incipient loading phase. {This second elementary mechanism of area variations will be named ``contact laying'' (or ``laying''). Finally, the third elementary mechanism is related to inhomogeneous slip and in-plane deformation within the contact zone and leads to either ``in-plane compression'' or ``in-plane dilation''. Such deformations extend beyond the contact region, which explains why the boundary of the green zone does not coincide with the dashed circle.}

\begin{figure}[ht!]
    \centerline{
    \inclps{1.0\textwidth}{!}{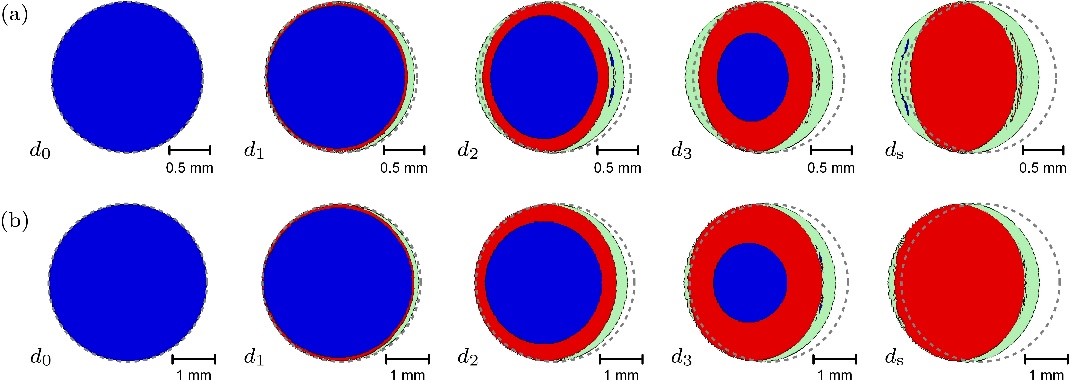}
    }
    \caption{Evolution of the contact zones for (a) $P=0.27$\,N ($d_1$, $d_2$, $d_3$: 35, 61, 81\% of $d_{\rm s}=0.45$\,mm) and (b) $P=2.12$\,N ($d_1$, $d_2$, $d_3$: 29, 58, 80\% of $d_{\rm s}=1.04$\,mm) shown in the frame attached to the moving rigid plate. Dashed circles indicate the boundary of the initial contact zone (at $d_0$). Blue and red regions denote the stick and slip zones, respectively. Green regions indicate the current (deformed) location of the initial contact zone that has been lifted. {The leading edge is on the left.}}
    \label{fig:evolution of contact regions glass-plate configuration}
\end{figure}

The existence of micro-slip in the model can be unambiguously identified in Fig.~\ref{fig:evolution of contact regions glass-plate configuration}, where the evolution of the contact during incipient loading is shown at five select tangential displacements of the rigid {plate:} $d_0=0$, $d_{\rm s}$ at the onset of gross sliding and three intermediate configurations ($d_i$). The nodes of the elastic surface are labeled in red if they have undergone some local slip, and in blue if they are still stuck to the same point of the rigid plate. It can be seen in Fig.~\ref{fig:evolution of contact regions glass-plate configuration} that the slip zone advances from the contact periphery towards the centre at the expense of the stick zone until the stick zone vanishes and full sliding occurs. As also observed in the experiments~\citep{sahli_shear-induced_2019}, the contact zone and the stick zone have an elliptical shape, the related shear-induced anisotropy increases with increasing displacement $d$ and is higher for the lower normal loads.

\begin{figure}[ht!]
    \centerline{
    \inclps{1.0\textwidth}{!}{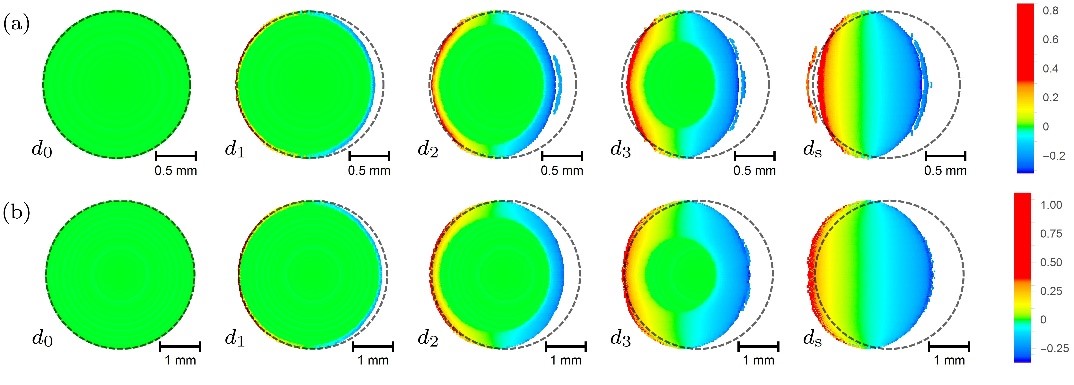}
    }
    \caption{Evolution of the field of {surface dilation/compression}, $(A^i-A^i_0)/A^i_0$, shown in the frame attached to the moving rigid plate for (a) $P=0.27$\,N and (b) $P=2.12$\,N, and for the same select displacements as in Fig.~\ref{fig:evolution of contact regions glass-plate configuration}. Dashed circles indicate the boundary of the initial contact zone. {The leading edge is on the left.}}
    \label{fig: In-plane area change glass}
\end{figure}

Partial slip configurations as those seen in Fig.~\ref{fig:evolution of contact regions glass-plate configuration} are classically found {(with a circular stick zone)} in models of sheared frictional linear elastic sphere/plane contact~{\citep{johnson_contact_1985,barber_contact_2018}}. They typically correspond to heterogeneous slip fields, thus causing in-plane deformations. Figure~\ref{fig: In-plane area change glass} provides a detailed insight into this mechanism, by showing the field of local {surface dilation/compression}. For each contact node $i$, its tributary area {(i.e.\ the area associated with the node, see also Section~S.1.1) is computed} both in the initial configuration (at $d_0$), $A^i_0$, and in the current configuration, $A^i$. The color map in Fig.~\ref{fig: In-plane area change glass} corresponds to the {local surface dilation/compression}, $(A^i-A^i_0)/A^i_0$. 
{Surface} dilation is observed close to the leading edge while compression is observed close to the trailing edge, which is consistent with the distribution of the $\sigma_{xx}$ stress component shown in Fig.~\ref{fig:3d fem meshes zoomed}(c). The decreasing green elliptical region in the middle of the contact zone, in which the {local surface dilation/compression} is equal to {zero}, corresponds to the stick zone in Fig.~\ref{fig:evolution of contact regions glass-plate configuration}.

\begin{figure}[th!]
    \centerline{
    \begin{tabular}{ccc}
    \inclps{0.40\textwidth}{!}{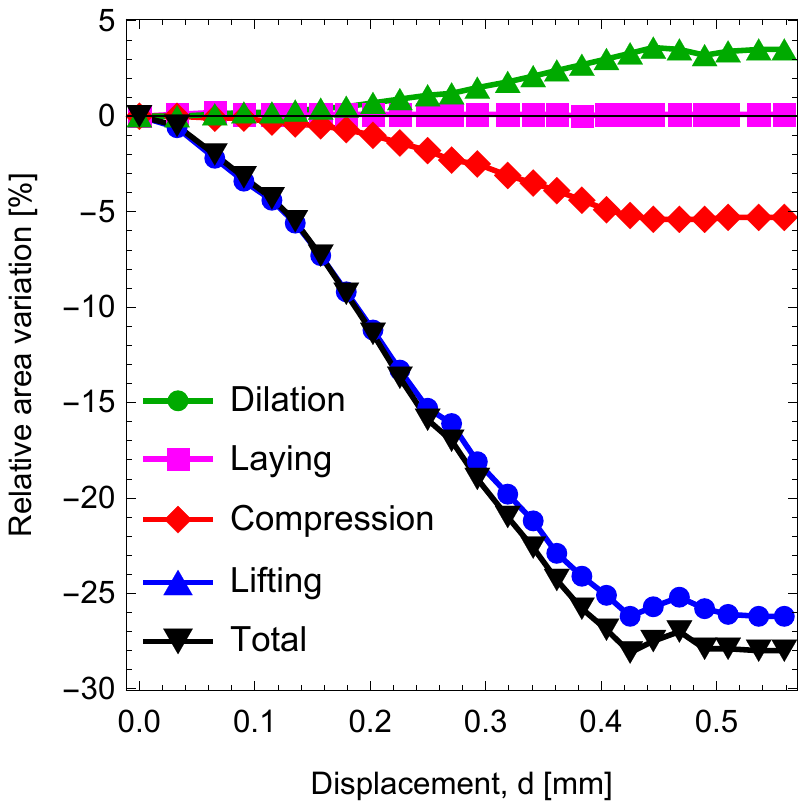} &&
    \inclps{0.40\textwidth}{!}{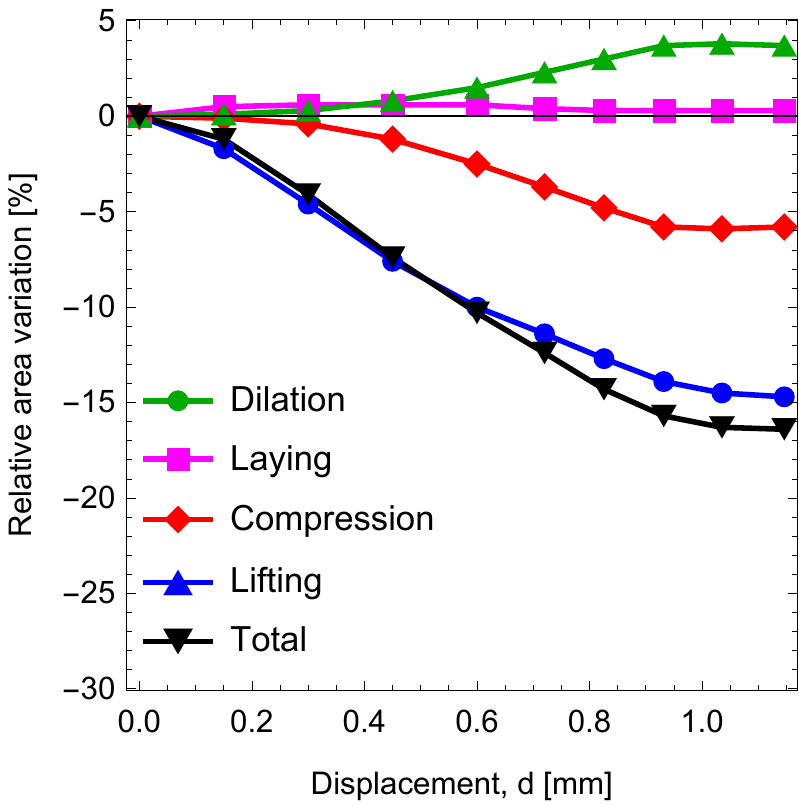} \\[0.5ex]
    \hspace*{2em}{\footnotesize (a)} & & 
    \hspace*{2em}{\footnotesize (b)}
    \end{tabular}
    }
    \caption{Evolution of the individual contributions of the various elementary mechanisms to the total relative area variation, $\Delta A/A_0$, as a function of the rigid plate displacement, $d$, for (a) $P=0.27$\,N and (b) $P=2.12$\,N.}
    \label{fig:Area reduction split plot LS model}
\end{figure}

{Figure~\ref{fig:Area reduction split plot LS model} compares the individual contributions of all mechanisms (calculated as described in \ref{app:fe-model}) to the total relative contact area change in the neo-Hookean model.} For all normal loads, the model predicts that lifting represents about 90--95\% of the total shear-induced area reduction, and is thus the primary mechanism responsible for it. The contributions of dilation and compression are individually significant, however, they essentially cancel out, so that the in-plane deformation is responsible only for about 5--10\% of the total area reduction. Contact laying is rarely observed (if so, at the leading edge only), so that its contribution to the total area reduction is essentially negligible in the model. 

{We emphasize that the relevant lifting in our neo-Hookean model is a local lifting due to shear deformations of the elastic sample, rather than a shear-induced global lifting of the rigid plate. Such a global lifting, although existing in our constant normal load setting, has been found to be only about 4\% of the normal displacement 
associated with the initial, purely normal loading. Thus, it is far from sufficient to explain the 15--30\% area reduction observed in Fig.~\ref{fig:Area reduction split plot LS model}.}

\subsection{{Qualitative role of finite deformations: half-space loaded by a tangential traction}}\label{sec:qualitative}

{To reach a general understanding of how finite deformations enhance anisotropic contact area reduction with respect to the linear elastic case, we studied a simplified auxiliary problem: a nearly incompressible hyperelastic half-space loaded by a homogeneous tangential traction $q_x$ (to mimick our Tresca model), applied over a circular area of radius $r$. In the finite-element model, the half-space is truncated, and the computational domain is a cylinder of radius $20r$ and height $20r$. The displacements at the truncation boundary are prescribed to those of the Boussinesq--Cerruti solution for a linear-elastic half-space loaded by a concentrated tangential force \citep{johnson_contact_1985}.}

\begin{figure}[ht!]
    \centerline{
    \begin{tabular}{ccc}
      \inclps{0.37\textwidth}{!}{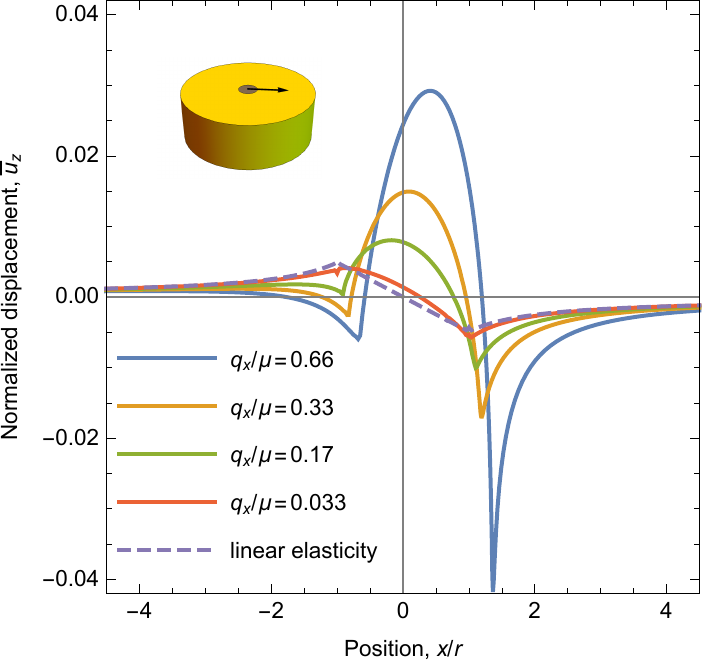}~ &
      \inclps{0.33\textwidth}{!}{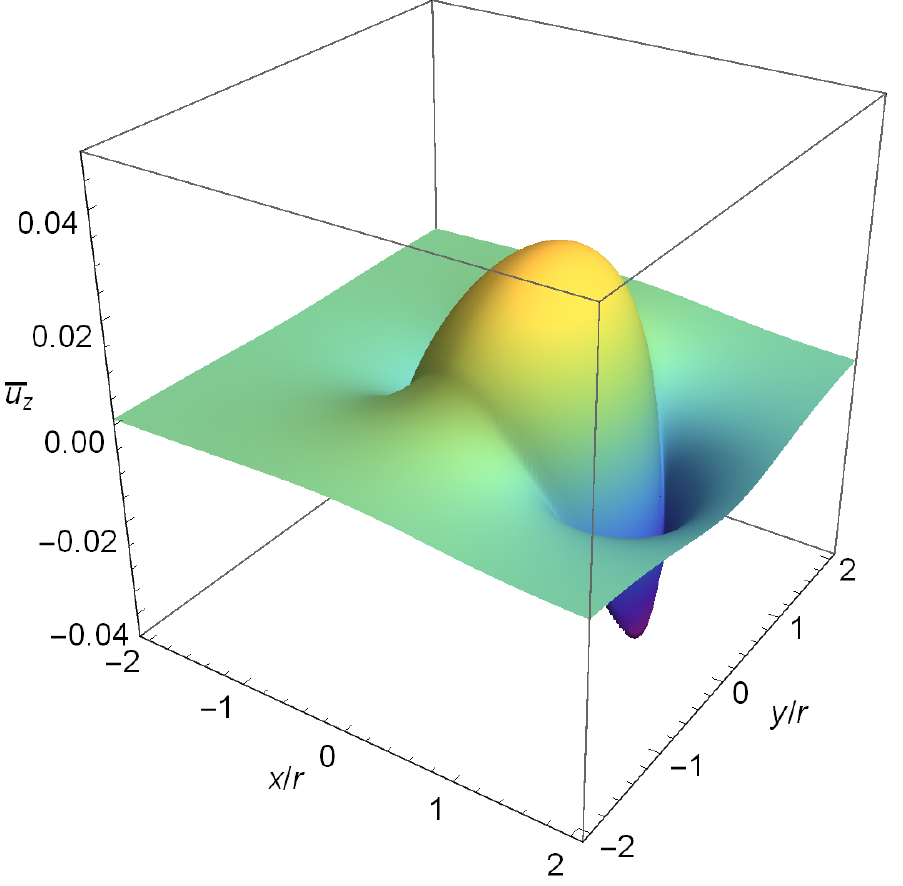} &
      \!\!\inclps{0.33\textwidth}{!}{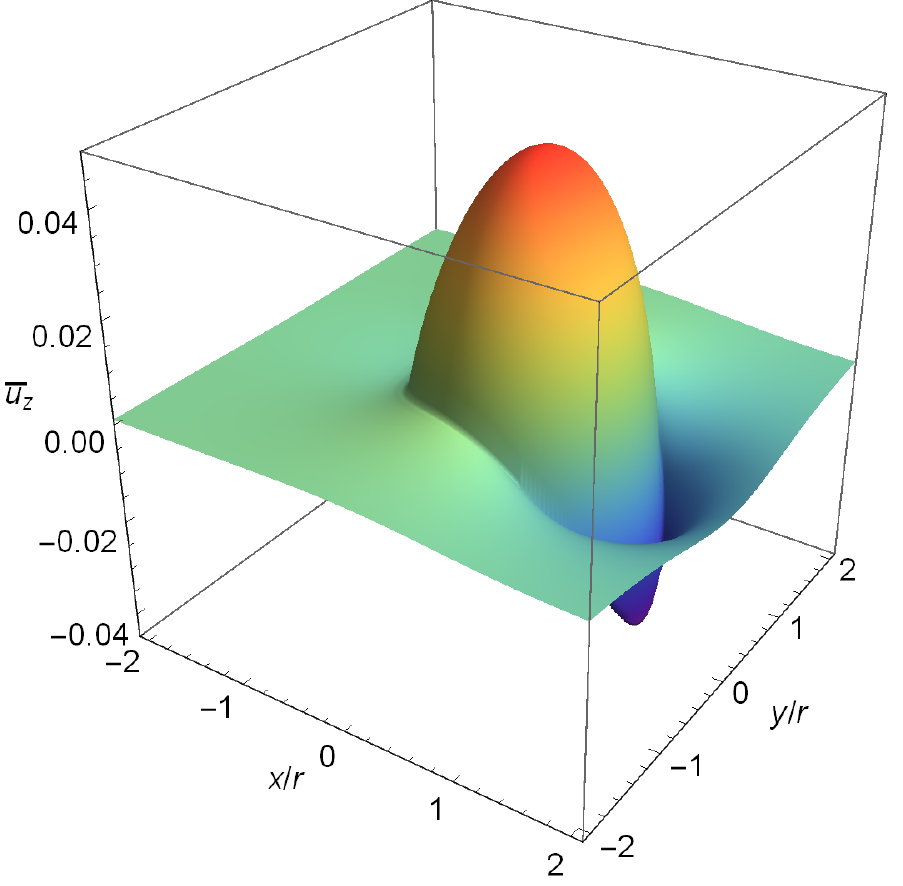}\!\! \\
      \hspace*{2em} {\footnotesize (a)} & {\footnotesize (b)} & {\footnotesize (c)}
    \end{tabular}
    }
    \caption{{Normalized vertical surface displacement, $\bar{u}_z=u_z\mu/(q_x r)$, for a nearly incompressible ($\nu=0.49$) half-space loaded by a homogeneous tangential traction $q_x$ along the $x$-axis, over a circular area of radius $r$. (a) Profile along the symmetry plane. Solid lines: neo-Hookean model for various  $q_x/\mu$. Dashed line: ($q_x$-independent) linear elastic case. (b,c) 2D maps within the $(x,y)$-plane for $q_x/\mu=0.66$: (b) neo-Hookean and (c) Mooney--Rivlin model. Spatial coordinates $(x,y)$ refer to the current configuration.}}
    \label{fig:halfspace}
\end{figure}

{Figure~\ref{fig:halfspace}(a) shows the normal surface displacements induced by the tangential loading, for both the linear elastic and neo-Hookean cases. Those displacements are the signature of a normal/tangential coupling.
A non-vanishing coupling is found in the linear elastic case because the material is only nearly incompressible. Still, the coupling remains weak, explaining why a significant contact area reduction is not found for linear elasticity. In the neo-Hookean case, for shear stresses  similar to those used in the contact problem ($q_x/\mu=0.66$ corresponds to $q_x=0.4$\,MPa, close to the value of $\sigma$ used in our Tresca model), the normalized displacements are much larger, indicating a strong normal/tangential coupling. The symmetry, which is characteristic for the linear-elastic model, is broken due to the finite-deformation effects. In particular, large negative displacements occur at the trailing edge, explaining why lifting has been found significant and localized at the trailing edge of the contact. Finally, the 2D map of the displacement field in Fig.~\ref{fig:halfspace}(b) clearly shows positive displacements in the direction orthogonal to shear. Such positive displacements found in the neo-Hookean case (resp.\ almost absent in the Mooney--Rivlin case, Fig.~\ref{fig:halfspace}(c)) are presumably responsible for the absence of lifting in the transverse direction, and thus for the contact anisotropy seen in Fig.~\ref{fig:contact zone different models 0.27}(b) (resp.\ absent in Fig.~\ref{fig:contact zone different models 0.27}(c)).}

\section{Illustration experiment}
\label{sec:experiment}

In this section, we perform an illustration experiment, similar to those of~\citet{sahli_evolution_2018,sahli_shear-induced_2019} used to validate the model in the previous section, in order to test whether the elementary mechanisms responsible for contact area reduction in the model (lifting, laying and in-plane deformation) can also be identified experimentally. The strategy is to incorporate particles close to the surface of the elastomer sphere and use them as tracers of the local motion of the frictional interface during incipient tangential loading.

\subsection{Experimental methods}\label{sec:expmethods}

\begin{figure}[ht!]
\centering
\includegraphics[width=0.72\linewidth]{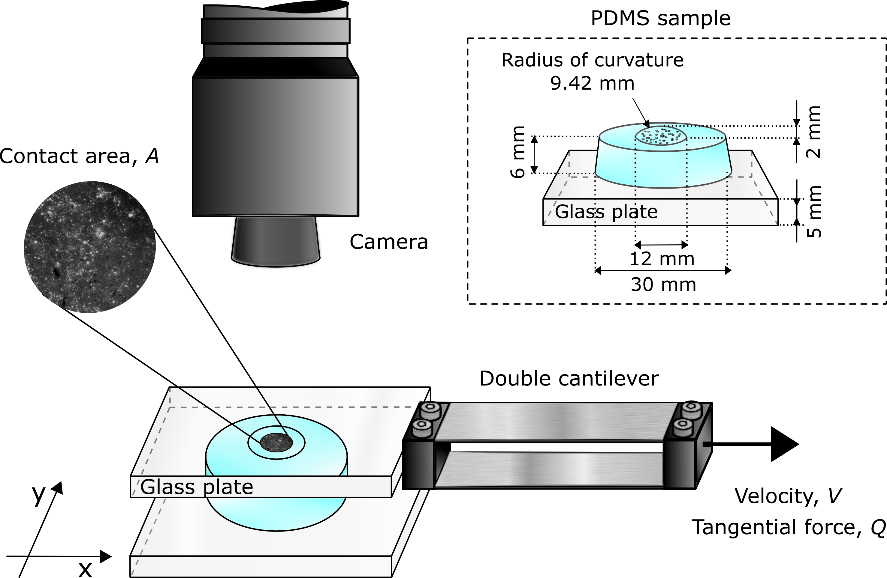}
\caption{Sketch of the opto-mechanical setup. The lower glass plate supporting the elastomer sample (inset) is attached to an optical table, while the substrate (upper glass plate) is driven tangentially at constant velocity $V$, through a horizontal double cantilever. The tangential force along $x$ is measured at the right extremity of the cantilever. The contact interface is illuminated from the top (lighting system not shown) and imaged in reflection with the camera. Inset: Sketch of the {tracer-seeded, spherically-topped} elastomer sample (blue, top).}
\label{fig:Setup_scretch}
\end{figure}

\begin{figure}[ht!]
\centering
\includegraphics[width=0.5\textwidth]{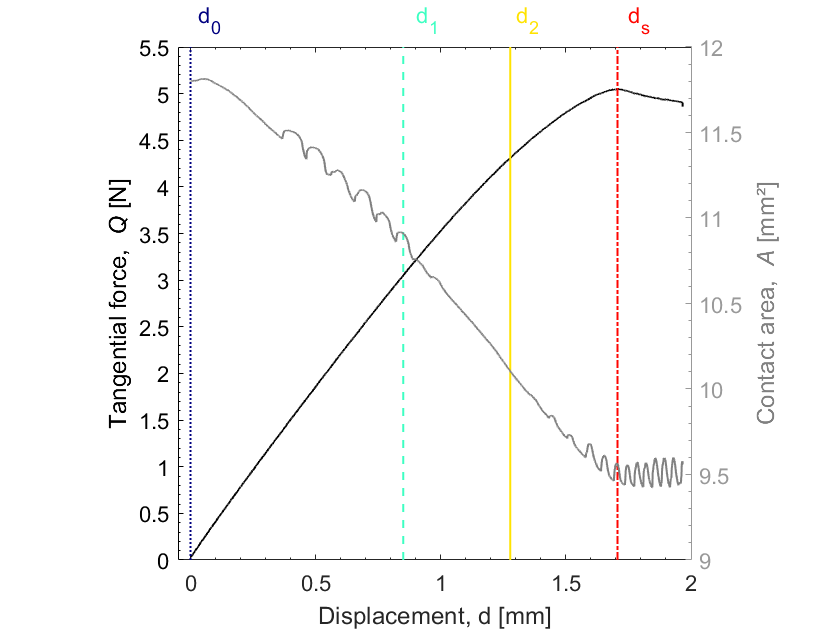}
\caption{Concurrent evolution of the tangential force $Q$ (solid black curve, left axis) and of the contact area $A$ (solid grey curve, right axis) as a function of the imposed tangential displacement of the glass substrate. $P=1.85$\,N. $V=0.1$\,mm/s. Four values of the displacement are indicated, which will be used in further {figures}.
}
\label{fig:Q_RCA_displacement}
\end{figure}

The illustration experiment was performed on a laboratory-built experimental setup ({see details in Section~S.2.1}) adapted from those used in~\cite{sahli_evolution_2018,sahli_shear-induced_2019,papangelo_shear-induced_2019,mergel_continuum_2019}, and sketched in Fig.~\ref{fig:Setup_scretch}. The two main improvements consisted in (i) replacing the single beam cantilever with a double beam cantilever and (ii) adding force sensors to measure the normal load. This setup is used to shear the interface between a smooth glass plate and a cross-linked PDMS sample {(nearly incompressible, $E = 1.5 \pm 0.1$\,MPa)} with a smooth spherical cap {seeded with a layer of particles about 16\,$\mu$m below the surface} ({see details in Section S.2.2}). 
The experiment presented here was performed under constant normal force $P=1.85$\,N. 19\,s after the contact has been created, a constant driving velocity $V=0.1$\,mm/s was imposed to the glass, over a total distance of 2\,mm. The evolution of the tangential force $Q$ as a function of the displacement of the glass plate is shown in black in Fig.~\ref{fig:Q_RCA_displacement}. The incipient tangential loading of the interface reaches a maximum, denoted as the static friction force $Q_{\text{s}}$, before a full sliding regime during which $Q<Q_{\text{s}}$. 

\begin{figure}[htb!]
\centering
\includegraphics[width=1\textwidth]{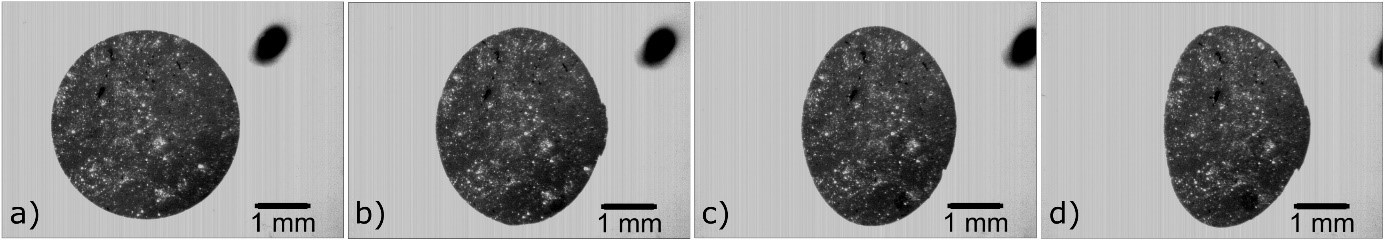}
\includegraphics[width=1\textwidth]{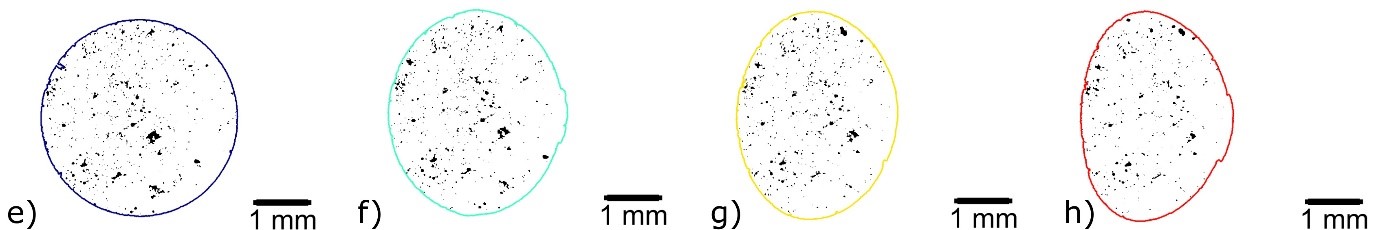}
\caption{Image analysis. Top row: raw images of the contact interface at the four displacements shown in Fig.~\ref{fig:Q_RCA_displacement}. Bottom row: corresponding segmented images. (a, e): $d_0$, (b, f): $d_1$, (c, g): $d_2$, (d, h): $d_{\text{s}}$. In (a-d), the main dark region is the contact, bright spots are the particles, and the {top right dark} region is a tracer drawn on the glass substrate {to monitor its macroscopic motion}. In (e-h), the contours are those of the contact (same colors as in Fig.~\ref{fig:Q_RCA_displacement}), the inner area of which defines the contact area; black spots correspond to the particles and are the {tracked} objects. {Images} (a-d) are in the frame of the camera, while (e-h) are in the frame of the (moving) glass plate. {The leading edge is on the left.}}
\label{fig:Images}
\end{figure}

Typical raw images of the contact interface are shown in Fig.~\ref{fig:Images}(a--d) for four different displacements (already shown in Fig.~\ref{fig:Q_RCA_displacement}): $d_0$ before any shear, $d_{\text{s}}$ the displacement at the static friction peak $Q_{\text{s}}$, and two intermediate displacements $d_1$ and $d_2$. The contact region corresponds to the biggest region of dark {pixels}. Figures~\ref{fig:Images}(e--h) show typical contours of the contact, the inner area of which defines the contact area, $A$. The evolution of $A$ as a function of the displacement imposed to the glass plate is shown in Fig.~\ref{fig:Q_RCA_displacement} (grey curve). Note that both curves in Fig.~\ref{fig:Q_RCA_displacement} show the same qualitative behaviour as that in previous experiments in the literature~\citep{waters_mode-mixity-dependent_2010,sahli_evolution_2018,mergel_continuum_2019}, indicating that the introduction of the particles into the elastomer sphere is not significantly affecting the mechanical response of the interface. The oscillations observable in the $A(d)$ curve in Fig.~\ref{fig:Q_RCA_displacement} originate from so-called re-attachment folds~\citep{barquins_sliding_1985} occurring at the trailing edge of the {contact}.

The contact region is not uniformly dark, {as in} our previous sphere/plane experiments~\citep{sahli_evolution_2018,sahli_shear-induced_2019,papangelo_shear-induced_2019,mergel_continuum_2019}, but is sprinkled with random bright spots which are due to {light reflection} on the particles {introduced} close to the elastomer surface. By using a homemade tracking procedure ({see details in Section S.2.3}), we were able to use those spots as tracers of the contact evolution and to generate a dataset in which the {$x$- and $y$-positions of each tracer is} given for each {image/time-step}. Note that only the tracers at the vertical of the contact regions are visible, so that they will be used not only to measure in-plane displacements but also to look for lifting {and laying (see Sections~\ref{sec:explifting} and~\ref{sec:explaying} respectively)}.

\subsection{Results and analysis}

In this section, based on the results of the tracking procedure, we perform a thorough analysis of the tracers' behaviour (appearance/disappearance {and in}-plane motion) to test the existence, in the experiments, of the elementary mechanisms found to be responsible for contact area evolution in the model (see Section~\ref{sec:mechanisms}). We first demonstrate that contact lifting and contact laying do occur, at opposite {sides of} the contact zone, and we quantify their amount using Voronoi tessellation. Second, using Delaunay triangulation, we illustrate the progressive development of a heterogeneous in-plane strain field within the contact area.

\subsubsection{Contact lifting}\label{sec:explifting}
{The fact that} only the particles at the vertical of the contact region can be {seen opens} the possibility to check whether lifting occurs at the interface. Indeed, the signature of local contact lifting is a particle that disappears from the image when reached by the moving contact periphery, meaning that a point of the elastomer initially in contact with the glass has been lifted out-of-contact. 

\begin{figure}[ht!]
\vspace*{-3ex}

\centerline{\includegraphics[width=0.55\linewidth]{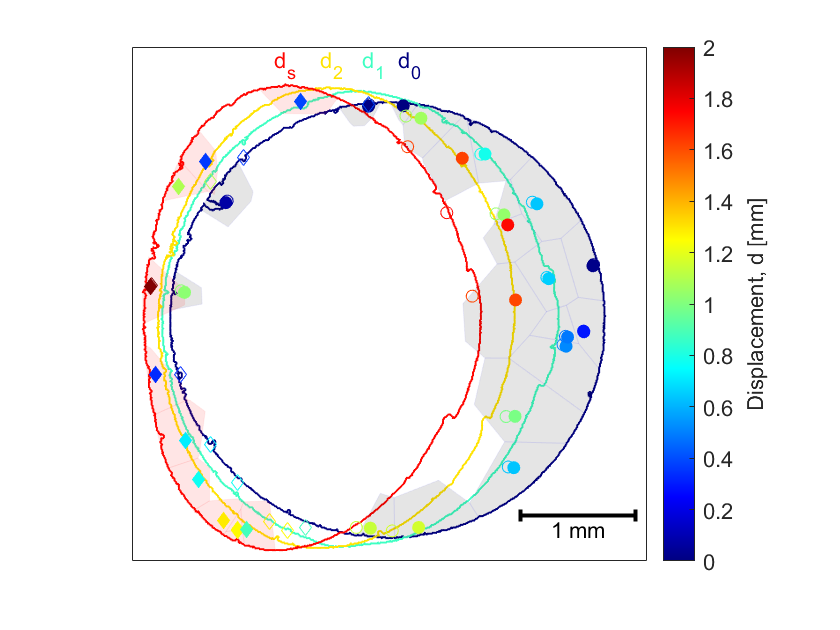}}
\vspace*{-4ex}
\caption{Measurement of the lifted and laid areas. Solid colored lines: contours of the contact at the four selected displacements (same colors as in Figs.~\ref{fig:Q_RCA_displacement} and~\ref{fig:Images}). Open (resp. filled) disks: position of the lifted particles when they disappear (resp. at $d_0$). The symbol color corresponds to the displacement at which the particle is lifted (same color code as for the contours). Grey cells: cells of the Voronoi tessellation at $d_0$ associated to lifted particles at $d_{\text{s}}$. Filled (resp. open) diamonds: position of the laid particles at $d_{\text{s}}$ (resp. when they appear). The symbol color corresponds to the displacement at which the particle are laid. Red cells: cells of the Voronoi tessellation at $d_{\text{s}}$ associated to laid particles at $d_{\text{s}}$. {The leading edge is on the left.}}
\label{fig:Peeling}
\end{figure}

In practice, we looked for tracer trajectories that ended at a location closer than 20 pixels to the contact periphery. {Such} tracers are represented {in Fig.~\ref{fig:Peeling}, either at their initial location (at $d_0$, filled disks) or where their disappear (open disks). A} significant number of tracers are indeed lifted during the incipient tangential loading of the contact. The large majority of them are found at the trailing edge of the {contact}, where points of the glass leave the contact. Two of them are found at the leading edge of the contact, but are probably tracers initially close to the contact periphery and disappearing due to noise in the images.
 
The color of each disk in Fig.~\ref{fig:Peeling} corresponds to the glass displacement at which the tracer disappears. It appears that the colors of the filled disks are spatially organized, from blue (early disappearance) for the rightmost disks to red (late disappearance) for the leftmost disks. Such a pattern indicates that lifting occurs through an inward front propagation, starting from the trailing {edge}.
 
\begin{figure}[ht!]
\centering
\includegraphics[width=0.5\textwidth]{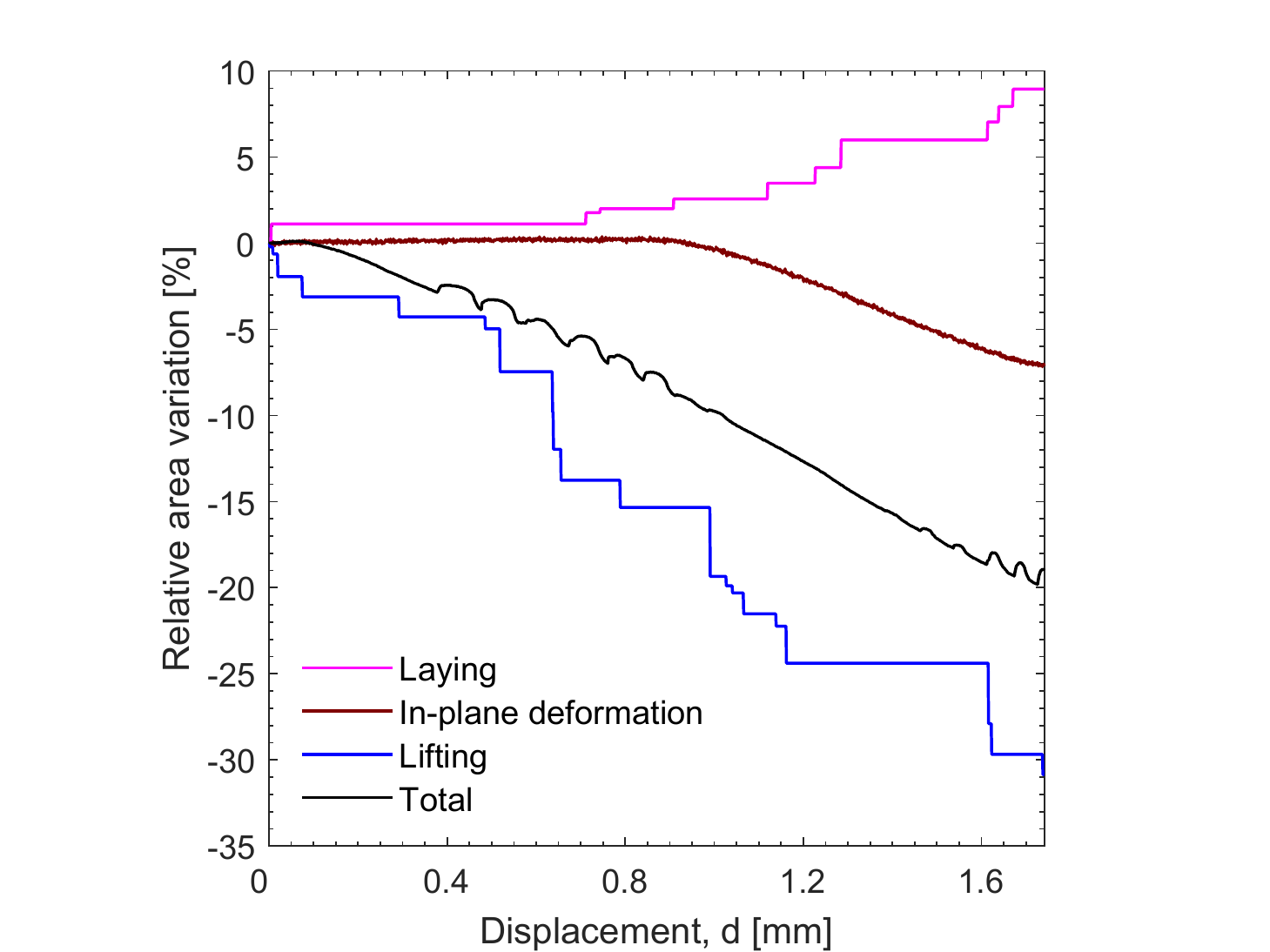}
\caption{Contributions of the various mechanisms to the area variation. {Relative area variation, $\Delta A / A_0$}, {vs} the imposed displacement, $d$. Black: reduction of the total contact area. Blue (resp. {pink}): area lost (resp. gained) by lifting (resp. laying), measured as in Fig.~\ref{fig:Peeling}. {Brown}: area variations due to slip-induced in-plane deformation, measured as in Fig.~\ref{fig:Deformation}.
}
\label{fig:Area_Voronoi}
\end{figure}

In order to quantify the amount of contact area that is lost due to lifting, we {assigned} a representative individual area to each tracer. For this, we performed, in the initial image, a bounded Voronoi tessellation on the centroids of the tracers, {the boundary being} the contact contour. Each tracer was thus assigned the area of its Voronoi cell. Then, {for each image}, the lifted area is computed as the sum of the areas of the Voronoi cells of all lifted tracers. The evolution of the lifted area along the experiment is shown in blue in Fig.~\ref{fig:Area_Voronoi}. The evolution is stair-like, because each lifting event increases abruptly the lifted area by a finite amount (the cell area). The estimated final lifted area represents about 31$\%$ of the initial contact area.
 
\subsubsection{Contact laying}\label{sec:explaying}
The mechanism of laying, i.e., points of the elastomer {getting} into contact with the glass upon shearing, can be analyzed in a similar way as lifting. We performed the same analysis {as} in the previous section, but backward in time: the first image considered {was the} one at the static friction peak (at $d_{\text{s}}$), while the last one was the initial image ({at $d_0$}). Doing so, tracers appearing close to the contact periphery in forward time are detected as disappearing in backward time.
 
The results of this analysis are shown in Fig.~\ref{fig:Peeling}{, where the locations of the laid tracers are shown either in the final configuration (at $d_{\text{s}}$, filled diamonds) or where they appear (open diamonds).} All laid tracers are found at the leading edge of the contact{, beyond the initial contact region, consistently with Fig.~\ref{fig:contact zone different models 0.27}a}. Their color, {evolving} from blue to red, corresponds to the stage at which they appear and is again spatially organized, indicating that contact laying occurs through an outward front propagation starting at the leading edge of the contact. The amount of contact area gained via laying is estimated using a bounded Voronoi tessellation performed in the final contact area (at $d_{\text{s}}$). {At each instant, the} laid area is counted as the sum of the areas of all laid tracers. In Fig.~\ref{fig:Peeling}, the red cells correspond to the area assigned to laying in the final image (at $d_{\text{s}}$). The evolution of the area gained thanks to {laying is} shown in red in {Fig.~\ref{fig:Area_Voronoi}. The} final amount of laid area is about 9$\%$ of the initial contact area. Unlike lifting, the laying  mechanism seems to initiate only after a finite shear has been applied.

\subsubsection{In-plane deformation}
Interestingly, in the previous analyses of both contact lifting and laying, the filled and open markers have different positions with respect to the glass substrate. This is a clear indication that slip occurs between the two instants: {local lifting} is preceded by slip, while {local laying} is followed by slip. Slip is found to be roughly parallel, but opposed to the glass motion, and occurs in both the leading and trailing regions of the contact.

{This observation is} consistent with a scenario of a micro-slip front propagating inward the contact region as shear is increased, which is classical in sheared sphere/plane contacts, either rough~\citep{prevost_probing_2013} or smooth~\citep{chateauminois_friction_2010}. In this scenario, which is also observed in the model results {of Section~\ref{sec:mechanisms}}, a peripheral slip region progressively invades the contact, replacing a shrinking central stuck region. From the combination of backward slip at the trailing edge and a stuck zone at the {contact center}, one {expects in-plane} compression of the {elastomer} in the trailing half of the contact. Symmetrically, in-plane dilation is expected in the leading half. The overall effect of both types of in-plane deformation must lead to a change in contact area that we {estimated as follows}.

\begin{figure}[ht!]
\centering
\includegraphics[width=1\linewidth]{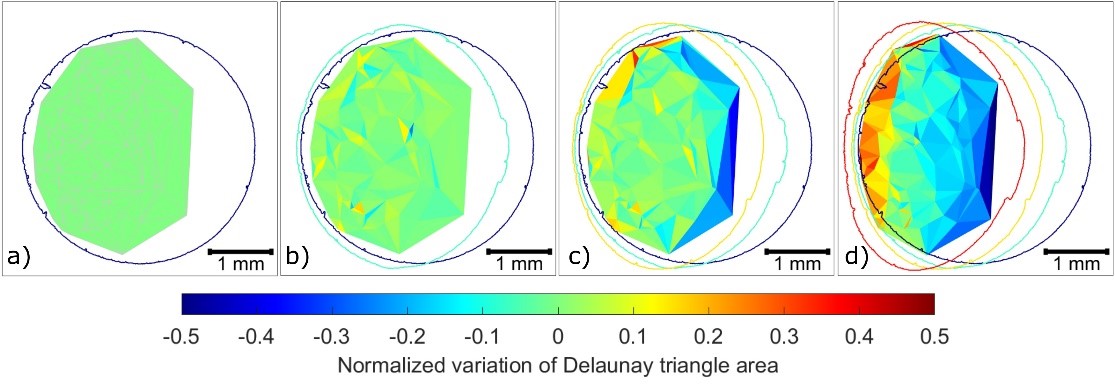}
\caption{Measurement {of in-plane} deformations. (a) Grey network: Delaunay tessellation based on all tracers {surviving from $d_0$ to $d_s$}. (b-d) Snapshots of the evolution of the Delaunay cells (defined at $d_0$) {at}: $d_1$ (b), $d_2$ (c) and $d_{\text{s}}$ (d). Solid lines are the contours of the contact at $d_0$, $d_1$, $d_2$ and $d_{\text{s}}$  with the same color code as in Figs.~\ref{fig:Q_RCA_displacement},~\ref{fig:Images}(e-h) and~\ref{fig:Peeling}. The color of each cell corresponds to its relative area change with respect to the situation at $d_0$ (see colorbar). Colder (resp. warmer) colors mean in-plane compression (resp. dilation).}
\label{fig:Deformation}
\end{figure}

{We} only considered the tracers present in the initial image and that could be followed during the whole experiment, from $d_0$ to $d_{\text{s}}$, thus excluding the lifted and laid ones. We performed a Delaunay triangulation to mesh all those tracers in the initial image, as shown as a gray network in Fig.~\ref{fig:Deformation}. As shear increases, the tracers move relative to each other. Keeping the same mesh, we follow the relative change of area of each {initial Delaunay triangles}, {as an estimator of local in-plane deformation.}

Fig.~\ref{fig:Deformation} shows the evolution of the field of in-plane deformation as shear is increased (cold colors for compression, warm ones for {dilation). Before} $d_1$ no deformation is observed, presumably because the slip front has not reached yet the fraction of initial contact covered by the Delaunay triangulation. Between $d_1$ and $d_2$, a compressed region sets in at the trailing side of the contact; the absence of a detectable symmetrical dilated region on the leading side suggests that the propagation of the slip front is not axisymmetrical, as usually considered in models of the incipient shear loading of circular {linear-elastic} contacts~\citep{barber_contact_2018}. At $d_{\text{s}}$, a heterogeneous {slip-induced} in-plane deformation field is fully developed, with a large compression region on the trailing side and a smaller dilation region on the leading side. Those field measurement {are qualitatively} consistent with previous measurements on a similar system, but made only along the central line of the contact in steady-state sliding~\citep{barquins_sliding_1985}.

The evolution under shear of the sum of {the total area} contained in the deforming Delaunay triangulation is shown in green in Fig.~\ref{fig:Area_Voronoi}. This area remains essentially unchanged during the first half of the experiment, and then progressively decreases by up to about 7$\%$ of the initial contact area.

\section{Discussion}

The model presented in Section~\ref{sec:model} has been shown to capture quantitatively the experimental results of~\cite{sahli_evolution_2018,sahli_shear-induced_2019} about the anisotropic shear-induced area reduction of sphere/plane elastomer contacts for relatively large normal loads (Fig.~\ref{fig:comparison}). We emphasize that this agreement is obtained without any adjustable parameter, in the sense that virtually all model parameters are set by independent measurements on the same experimental system. Those parameters are: {the geometry of the hyperelastic neo-Hookean sample, its shear} modulus, and the contact shear strength of the PDMS/glass interface. Note that the elastic bulk modulus (in our nearly incompressible framework), the mesh size and the timescale in the regularized Tresca model are three purely numerical parameters, the precise values of which have been checked to negligibly affect our results. In our opinion, such an absence of adjustable parameter provides our model with a decisive comparative advantage with respect to competing models of shear-induced contact area reduction. In particular, {fracture-like adhesion-based models}, in order to offer a good quantitative agreement with experiments~\citep{papangelo_shear-induced_2019,das_sliding_2020}, require prescription of {finely-tuned} mode-mixity functions~\citep{papangelo_mixed-mode_2019}. {Our 3D model also} naturally captures the anisotropic evolution of the contact shape, whereas most fracture-based adhesive models assume axisymmetry of the contact (see~\cite{papangelo_shear-induced_2019} for the only exception, to our knowledge).

The contact area behaviour of {our neo}-Hookean model has {been validated} using a competing finite-element model based on different regularization options, meshing and solver ({\ref{app:validation}}). This good agreement demonstrates the robustness of our model results to all implementation {details.} 

The direct implication of our model is {that finite}-deformation effects and the non-linear elasticity of elastomers are presumably the key ingredients explaining those experimental results, rather than {viscoelasticity or} adhesion. Nevertheless, our model has only been applied to {contacts submitted} to relatively large normal load, in the newton range~\citep{sahli_evolution_2018}, while other {experiments have used} much smaller normal loads, in the millinewton range~\citep{savkoor_effect_1977, waters_mode-mixity-dependent_2010,mergel_continuum_2019}. For those lighter loads, adhesive stresses may be of the order of, or even exceed, contact stresses, and non-linear elasticity may be not the {only} dominant ingredient for shear-induced contact shrinking anymore. {Identifying} the normal load regimes in which {adhesion needs to be accounted for} is an important goal for future contact mechanics models. In the following, we will restrict our discussion to the high-load regime {explored in the present} study.

By comparing models assuming either small or finite strains (Fig.~\ref{fig:area reduction}), we have shown that {finite deformations} is the single model ingredient responsible for the significant shear-induced contact area reduction observed in sphere/plane {elastic} contacts. {Although the amplitude of the reduction significantly depends on the particular hyperelastic model used (Fig.~\ref{fig:area reduction})}, {the respective roles of hyperelasticity and of the exact geometrically non-linear kinematics are difficult to disentangle. What is clear is that finite deformations induce a significant coupling of the normal and tangential deformation modes, which is lacking in the small-strain incompressible elasticity. Such a coupling generates substantial vertical displacements which account both for a significant lifting localized at the trailing edge and for the marked anisotropy of the final contact shape (Fig.~\ref{fig:halfspace}).}

{We quantified}, for the first time, the various elementary mechanisms by which area variations {can occur: contact lifting, contact laying} and in-plane deformation. Our model results suggest that, in the experimental conditions used in~\cite{sahli_evolution_2018,sahli_shear-induced_2019}, {local} lifting is the dominant mechanism explaining the observed significant area reduction and its anisotropy (Fig.~\ref{fig:Area reduction split plot LS model}). This conclusion {is true} for all normal loads explored, with the total reduction {amplitude being} dependent on the normal load (as also observed experimentally). In future works, it would be interesting to vary systematically all model parameters beyond the experimental {range, to} clarify their respective roles in the {area variations due to each of the three mechanisms}.

{The} incipient shear-loading of smooth sphere/plane elastic contacts is characterized in the model by two propagating fronts: a lifting front at the contact periphery, and a {non-circular micro-slip front within} the contact. The existence of those two different {fronts, although} explicitly acknowledged by some authors~\citep{johnson_adhesion_1997,mcmeeking_interaction_2020}, has never been properly described in fracture-based adhesive models. Indeed, {the additional energy dissipation in the contact due to frictional micro-slip, and described via the mode-mixity function,} is assumed to be located at the contact periphery, and not within a growing, finite region of the contact.

The relevance of {our} model observations has been qualitatively confirmed by the original experimental results of Section~\ref{sec:experiment}. Using an elastomer sphere seeded with tracers, we demonstrated that shear-induced anisotropic contact area reduction is indeed accompanied by all three possible mechanisms. Although a higher areal density of tracers would be desirable to reduce the measurement uncertainties (\ref{app:uncertainties}), we could conclude that lifting is presumably the dominant area reduction mechanism, also in the experiments. While experiments revealed a potentially larger contribution of the laying  and in-plane mechanisms, a definitive quantitative comparison with the model is hindered by the finite experimental resolution. Also note that, due to the higher contact shear strength (0.53 rather than 0.41\,MPa) and smaller Young's modulus (1.5 rather than 1.8\,MPa) of the present experiments compared to those of~\cite{sahli_evolution_2018,sahli_shear-induced_2019}, our numerical model could not converge in those more severe conditions, thus impeding direct comparison between model and tracer-based experiments. Such a comparison is also an important challenge for future works.

Finally, the simplicity and generality of our model assumptions suggest that our results may be also relevant to other systems than the elastomer sphere/plane contacts studied here. First, non-linear elasticity being a generic feature of soft materials, from gels to human skin, we expect it to be a likely mechanism for contact area reduction in all studies involving {such} materials. It would thus be interesting to re-interpret recent experiments like those of e.g.~\cite{das_sliding_2020} on polyacrylonitrile or those of~\cite{sahli_evolution_2018,sirin_fingerpad_2019} on human fingertips, from the standpoint of finite-deformation {mechanics. Second,~\cite{sahli_evolution_2018,sahli_shear-induced_2019} argued} that the mechanisms of shear-induced anisotropic contact area reduction may be the same in millimetric sphere/plane contacts and in individual micrometric junctions within rough contact interfaces. Such a similarity across scales suggests that the {present conclusions} may also be used to further understand the shear behaviour of soft material multicontacts, {the modelling of which may require non-linear elasticity}. This may in particular be an explanation for the fact that, in~\cite{scheibert_onset_2020}, a multi-asperity model based on linear elasticity failed to reproduce quantitatively the multicontact results of~\cite{sahli_evolution_2018}.

\section{Conclusion}

We have developed the first non-adhesive, non-viscous model of shear-induced contact area reduction in soft materials. Quantitative agreement with the recent experimental results of~\citet{sahli_evolution_2018,sahli_shear-induced_2019} on sphere/plane elastomer contacts has been obtained, without adjustable parameter, using the Tresca friction law and the neo-Hookean hyperelastic model. {The} necessary ingredients for this agreement are finite deformations and non-linear elasticity.

A detailed analysis of the model has revealed the elementary mechanisms responsible for contact area reduction. Local contact lifting {dominates}, while in-plane dilation and compression, although significant, approximately cancel each other. Those predicted mechanisms, and their relative contributions to area reduction, have been confirmed experimentally by tracking tracers introduced {in an elastomer/glass interface}, and applying an original analysis to the tracking data. The challenges associated with both the modelling and experimental approaches have been thoroughly discussed.

All our results suggest a new perspective on the phenomenon of shear-induced contact area reduction in soft materials. This currently highly debated topic has been dominated by interpretations based on a {leading} role of adhesion. Here, instead, we suggest that finite-deformation effects and non-linear elasticity can be equally important, all the more so as large normal loads are considered. Clarification of the {validity domains where adhesion and/or finite deformations need to be accounted for} remains as a major open issue on the topic.

\paragraph{Acknowledgement} This work was partially supported by the EU Horizon 2020 Marie Sklodowska Curie Individual Fellowship \emph{MOrPhEM} (H2020-MSCA-IF-2017, project no.\ 800150). DD, MdS and JS are indebted to Institut Carnot Ing\'enierie@Lyon for support and funding. They also thank Crist\'obal Oliver Vergara for his help in the initial development of the particle tracking procedure. {We thank an anonymous reviewer for suggesting the simulations described in Section~\ref{sec:qualitative}.}

\appendix

\section{Description of the finite-deformation model}
\label{app:model}

\setcounter{figure}{0}

\subsection{Finite-strain framework}
\label{Sec: finite strain}

{The finite-strain framework} is standard, so only the basic notions are provided below for completeness, with the scope limited to hyperelasticity. 
Two configurations of the body are considered: the reference configuration $\Omega$ and the current (deformed) configuration $\omega$. 
The deformation that brings $\Omega$ to $\omega$ is described by the deformation mapping $\bm{\varphi}$ such that $\bm{x}=\bm{\varphi}(\bm{X})$, where $\bm{X}\in\Omega$ and $\bm{x}\in\omega$ denote the position of a material point in the respective configuration. 
{The deformation gradient is defined as} $\bm{F}=\operatorname{Grad}\bm{\varphi}$, where $\operatorname{Grad}$ is the gradient with respect to the reference configuration $\Omega$.

Equilibrium equation in the strong form is written in the reference configuration $\Omega$, and, in the absence of body forces, reads
\begin{equation}
  \operatorname{Div} \bm{P} = \bm{0} , \qquad
  \bm{P} = \frac{\partial W}{\partial \bm{F}} ,
\end{equation}
where $\bm{P}$ is the first Piola--Kirchhoff stress tensor, $\operatorname{Div}$ is the divergence operator relative to the reference configuration $\Omega$, and the elastic strain energy $W=W(\bm{F})$ specifies the constitutive response of a hyperelastic body. 

The reference material model used in this work is the nearly-incompressible isotropic neo-Hookean model specified by $W=W_{\rm nH}$,
\begin{equation}
\label{Eq: NH model}
  W_{\rm nH}(\bm{F}) = \frac{1}{2} \mu (\bar{I}_1 - 3) + W_{\rm vol}(J) ,
  \quad
  W_{\rm vol}(J) = \frac{1}{4} \kappa \left( (J-1)^2 + (\log J)^2 \right) ,
\end{equation}
where $W_{\rm vol}(J)$ is the volumetric part of the elastic strain energy, $J=\det\bm{F}$, and $\bar{I}_1=\operatorname{tr}\bar{\bm{b}}$ is the invariant of $\bar{\bm{b}}=J^{-2/3}\bm{b}$, $\bm{b}=\bm{F}\bm{F}^{\rm T}$. 
Material properties are here specified by the shear modulus $\mu$ and bulk modulus $\kappa$. 

The neo-Hookean model is a special case of the Mooney--Rivlin model specified by $W=W_{\rm MR}$,
\begin{equation}
\label{Eq: MR model}
  W_{\rm MR}(\bm{F}) = \frac{1}{2} \mu_1 (\bar{I}_1 - 3) {}+ \frac{1}{2} \mu_2 (\bar{I}_2 - 3) + W_{\rm vol}(J) ,
\end{equation}
where $\bar{I}_2=\frac{1}{2}(\bar{I}_1^2-\operatorname{tr}\bar{\bm{b}}^2)$ and $\mu=\mu_1+\mu_2$. The neo-Hookean model is recovered for $\mu_2=0$. The other particular case of the Mooney--Rivlin model used in this study corresponds to $\mu_1=\mu_2=\mu/2$, and is simply called Mooney--Rivlin in all model results.

\subsection{Contact with a rigid plane}\label{sec:kinematics}

The contact formulation briefly described below is limited to the case of contact with a rigid plane, which is sufficient for the purpose of this work, see e.g.\ \citet{Wriggers06} for a more general presentation. 
The orientation of the plane is specified by the outward unit normal $\bm{\nu}$, {and the motion} of the rigid plane is restricted to a translation, thus $\dot{\bm{\nu}}=\bm{0}$. 

Contact kinematics is specified by the normal gap $g_{\rm N}$ and tangential slip velocity $\bm{v}_{\rm T}$ that are defined for each point $\bm{x}$ on the potential contact surface $\gamma_{\rm c}=\bm{\varphi}(\Gamma_{\rm c})$, where $\Gamma_{\rm c}$ denotes the contact surface in the reference configuration. {Here}, the kinematic quantities are simply given by
\begin{equation}
  \label{eq:gNvT}
  g_{\rm N} = (\bm{x}-\bm{x}_{\rm R}) \cdot \bm{\nu} , \qquad
  \bm{v}_{\rm T} = (\bm{1}-\bm{\nu}\otimes\bm{\nu}) (\dot{\bm{x}}-\dot{\bm{x}}_{\rm R}) ,
\end{equation}
where $\bm{x}_{\rm R}$ denotes the current position of a fixed point on the rigid plane, {and $\otimes$ denotes the diadic product.}

The contact traction vector $\bm{t}$ is defined as the traction vector exerted by the body on the rigid plane. 
By the action-reaction principle it {opposes the traction vector acting on the body, so} we have $\bm{t}=-\bm{\sigma}\bm{n}$, where $\bm{\sigma}$ is the Cauchy stress tensor, $\bm{n}$ is the outward unit normal to the contact surface $\gamma_{\rm c}$, and {$\bm{n}=-\bm{\nu}$ whenever contact occurs}. 
Note that $\bm{t}$ is a \emph{spatial} traction vector, i.e., it refers to a unit area in the \emph{current} configuration. 
{$\bm{t}$} is decomposed into its normal and tangential parts relative to the rigid-plane normal $\bm{\nu}$, namely
\begin{equation}
  t_{\rm N} = \bm{t} \cdot \bm{\nu} , \qquad
  \bm{t}_{\rm T} = (\bm{1}-\bm{\nu}\otimes\bm{\nu}) \bm{t} ,
\end{equation}
so that $\bm{t}=t_{\rm N}\bm{\nu}+\bm{t}_{\rm T}$.

With the {above} definitions, unilateral contact conditions {are expressed as}: \begin{equation}
  \label{eq:unilateral}
  g_{\rm N} \geq 0 , \qquad t_{\rm N} \leq 0 , \qquad g_{\rm N} t_{\rm N} = 0 .
\end{equation}
The friction model is discussed in \ref{sec:Tresca}.

{Finally}, the virtual work principle, i.e., the weak form of the mechanical equilibrium, which is the basis for the finite-element implementation, {is expressed as}:
\begin{equation}
  \label{eq:VWP}
  \int_\Omega \bm{P} \cdot \operatorname{Grad}\delta\bm{\varphi} \, {\rm d} V + 
  \int_{\gamma_{\rm c}} ( t_{\rm N} \delta g_{\rm N} + \bm{t}_{\rm T} \cdot \delta \bm{g}_{\rm T} ) {\rm d} s = 0 \qquad \forall \, \delta \bm{\varphi} ,
\end{equation}
where $\delta\bm{\varphi}$ is the test function that vanishes on the part of the boundary of $\Omega$ on which the displacement is prescribed, and we have $\delta g_{\rm N}=\bm{\nu}\cdot\delta\bm{\varphi}$ and $\delta\bm{g}_{\rm T}=(\bm{1}-\bm{\nu}\otimes\bm{\nu})\delta\bm{\varphi}$ (cf.\ Eq.~\eqref{eq:gNvT}). 
Since $t_{\rm N}$ and $\bm{t}_{\rm T}$ are spatial tractions, the contact contribution, i.e., the second term in Eq.~\eqref{eq:VWP}, is integrated over the contact surface $\gamma_{\rm c}$ in the current configuration.

\subsection{Regularized Tresca friction model}
\label{sec:Tresca}

\begin{figure}[hbt]
    \centering
    \subfloat[]{\inclps{0.23\textwidth}{!}{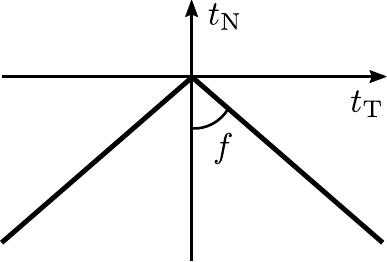}} \hspace{0.04\textwidth}
    \subfloat[]{\inclps{0.23\textwidth}{!}{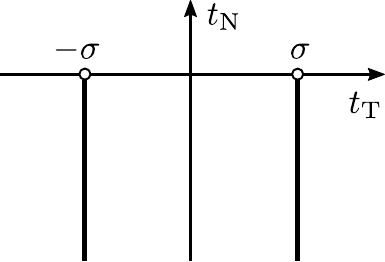}} \hspace{0.04\textwidth}
    \subfloat[]{\inclps{0.23\textwidth}{!}{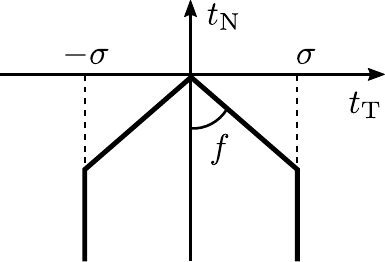}}
    \caption{Sketch of three friction laws: (a) Coulomb with a friction coefficient $f$, (b) Tresca with a contact shear strength $\sigma$ and (c) Coulomb--Orowan with both a friction coefficient $f$ and a contact shear strength $\sigma$.}
    \label{fig:tresca}
\end{figure}

In the Tresca friction model, the limit friction stress, called here the contact shear strength, $\sigma$, is constant and independent of the normal contact traction $t_{\rm N}$ (Fig.~\ref{fig:tresca}(b)). 
The Tresca model can be written {as}:
\begin{equation}
  \label{eq:tresca}
  \|\bm{t}_{\text{T}}\|-\sigma \leq 0 , \qquad
  \|\bm{t}_{\text{T}}\| \bm{v}_{\text{T}} = \bm{t}_{\text{T}} \|\bm{v}_{\text{T}}\| , \qquad
  ( \|\bm{t}_{\text{T}}\|-\sigma ) \|\bm{v}_{\text{T}}\| = 0 ,
\end{equation}
and the equations above hold in the case of \emph{active contact}, i.e., for $g_{\rm N}=0$. 
In the case of \emph{separation}, i.e., for $g_{\rm N}>0$, the tangential traction vanishes, $\bm{t}_{\rm T}=\bm{0}$. 
Note that, in view of the unilateral contact condition (Eq.~\eqref{eq:unilateral}), penetration ($g_{\rm N}<0$) is not allowed.

{The first condition in Eq.~\eqref{eq:tresca} is the limit friction {condition.}} The second condition is the slip rule that implies that the direction of the tangential (slip) velocity $\bm{v}_{\rm T}$ is that of the tangential traction $\bm{t}_{\rm T}$. {The} third (complementarity) condition controls the stick/slip state such that sticking contact ($\bm{v}_{\rm T}=\bm{0}$) occurs when $\|\bm{t}_{\text{T}}\|<\sigma$ and sliding ($\|\bm{v}_{\rm T}\|>0$) may occur only when $\|\bm{t}_{\text{T}}\|=\sigma$.

{One feature of the Tresca model} is that the tangential traction $\bm{t}_{\rm T}$ may suffer a jump at the instant of transition between active contact and separation. This, in turn, may lead to significant convergence problems in the respective computational scheme, for instance, in the framework of the finite-element method. Accordingly, a regularization scheme has been developed, as described below, and employed in the actual computations. Note that an alternative approach, based on the Coulomb--Orowan friction model (Fig.~\ref{fig:tresca}(c)), has also been examined, and the respective results are briefly presented in {\ref{app:validation}, see also Section~S.1.3. The} Tresca model {is the} limit case of the Coulomb--Orowan model for $f\to+\infty$.

In the regularization scheme adopted here, it is assumed that the tangential traction $\bm{t}_{\rm T}$ does not drop to zero instantly after separation occurs but requires some characteristic time $\tau$ to gradually vanish. Accordingly, the following simple evolution law is adopted in the case of \emph{separation} ($g_{\text{N}}>0$):
\begin{equation}
  \label{eq:regularization}
  \dot{\bm{t}}_{\text{T}}=
  \begin{cases}
    \displaystyle
    -\frac{\sigma}{\tau} \frac{\bm{t}_{\text{T}}}{\|\bm{t}_{\text{T}}\|} \qquad
      & \text{for} \;\; \|\bm{t}_{\text{T}}\|>0 , \\
    \bm{0} & \text{otherwise},
  \end{cases}
\end{equation}
so that the magnitude of the tangential traction $\bm{t}_{\rm T}$ decreases towards zero at a constant rate of $\sigma/\tau$. A similar regularization, leading to a kind of rate-and-state friction law, has been adopted by \citet{CochardRice00} to regularize the abrupt changes in friction associated with abrupt changes in the normal contact traction, see also \citet{PrakashClifton1993, Prakash1998} for the experimental justification and respective constitutive modelling. 
Here, the regularization is introduced purely for computational reasons.

\subsection{Finite-element model}
\label{app:fe-model}

{Finite-element implementation has been performed using the \emph{AceGen/AceFEM} system \citep{KorelcWriggers16}. Eight-node hexahedral TSCG12 elements \citep{Korelc10} are used for the solid. 
The augmented Lagrangian method \citep{AlartCurnier91} combined with nodal integration is used to enforce the unilateral contact and friction conditions (Eqs.~\eqref{eq:unilateral} and~\eqref{eq:tresca}), see \citet{LengKorStu11} for the respective automation and \emph{AceGen}-based implementation. More details are provided in Section~S.1.1.}

The geometric parameters and material properties used in the simulations are directly taken from one of the experimental datasets reported in~\citet{sahli_evolution_2018} (smooth PDMS-sphere/bare-glass contact, see Fig.~2C therein). 
The boundary conditions and the loading sequence also correspond to those in the experiment. The displacements are fully constrained at the bottom surface of the elastomer sample (cf.\ Figs.~\ref{fig: fem schematics} and \ref{fig:Setup_scretch}), except for a rigid-body translation in the $z$-direction so that the normal load $P$ can be applied against a rigid plane representing the upper glass plate. Subsequently, a constant tangential velocity of the rigid plane is prescribed along the $x$-direction while the normal load $P$ is kept constant. Further, the symmetry with respect to the $y=0$ plane is exploited so that only one half of the sample is effectively modelled with adequate boundary condition imposed on the symmetry plane.

The resulting finite-element mesh is shown in Fig.~\ref{fig:3d fem mesh}. The sample is discretized with about 100,000 hexahedral elements, which gives approximately 380,000 degrees of freedom. 
The hanging-node technique is used to conveniently refine the mesh in the vicinity of the potential contact surface, and the size of the refined-mesh region is adjusted to the normal load. The contact surface comprises about 12,600 nodes with the size of the quadrilateral elements ranging from 13\,$\mu$m for $P=0.27$\,N to 25\,$\mu$m for $P=2.12$\,N. 

The elastic properties of the elastomer sample have been identified by matching the contact area obtained from the computational model with neo-Hookean elasticity for the highest normal load ($P=2.12$\,N) and zero tangential load ($Q=0$) to that measured in the experiment. Assuming a nearly incompressible response with the Poisson's ratio $\nu=0.49$, the Young's modulus has been identified as $E=1.80$\,MPa, which {is fully consistent with the value provided in~\citet{sahli_evolution_2018}, and} corresponds to a shear modulus $\mu=0.60$\,MPa. 
The relationship between the parameters $(E, \nu)$ and $(\mu, \kappa)$ (cf.\ Eq.~\eqref{Eq: NH model}) is given by $\mu=E/(2(1+\nu))$ and $\kappa=E/(3(1-2\nu))$. It can be seen in Fig.~\ref{fig:comparison} that, for the shear modulus determined as described above, the contact area at zero tangential load is correctly represented in the whole range of normal loads for both the neo-Hookean and Mooney--Rivlin models. {The fact that, at the smallest loads, the models slightly underestimate the initial contact area is presumably due to the non vanishing effect of adhesion in the experiments.}

The value of the contact shear strength $\sigma=0.41$\,MPa in the Tresca model (Eq.~\eqref{eq:tresca}) is taken as the slope of the best linear fit of the evolution of the static friction force, $Q_{\rm s}$ as a function of the concurrent contact area, $A_{\rm s}=A(Q=Q_{\rm s})$ in the experimental {dataset.}

The Prakash-Clifton-like regularization parameter in the Tresca model is assumed to be $\tau=0.2$\,s, a value which has been checked to be small enough to have negligible influence on the results. 
The computations have been carried out using an adaptive time stepping procedure. The typical number of resulting time increments was between 50 and 100, which corresponds to an average time increment of the order of 0.1\,s. Note that the data points included in the figures reported throughout Section~\ref{sec:model} and \ref{app:model} correspond to selected instants.

\begin{figure}[htb!]
    \centerline{
    \hspace*{0.\textwidth}
    \begin{tabular}{ccc}
    \inclps{0.25\textwidth}{!}{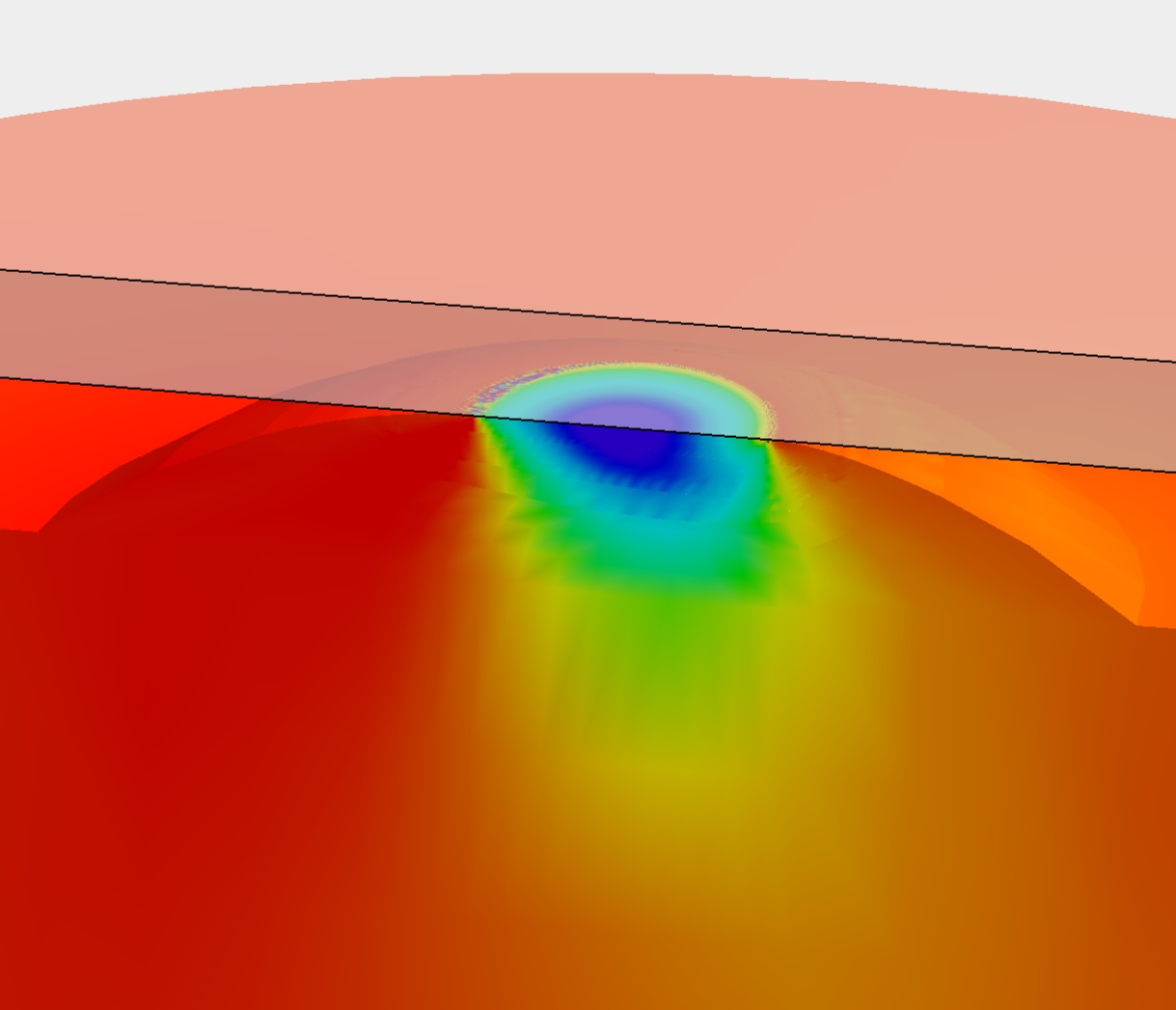}
    \inclps{0.06\textwidth}{!}{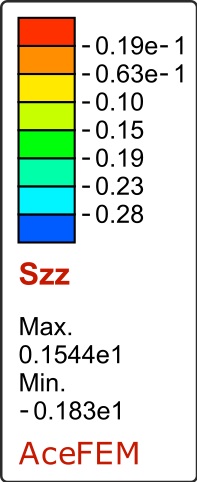}
    &
    \inclps{0.25\textwidth}{!}{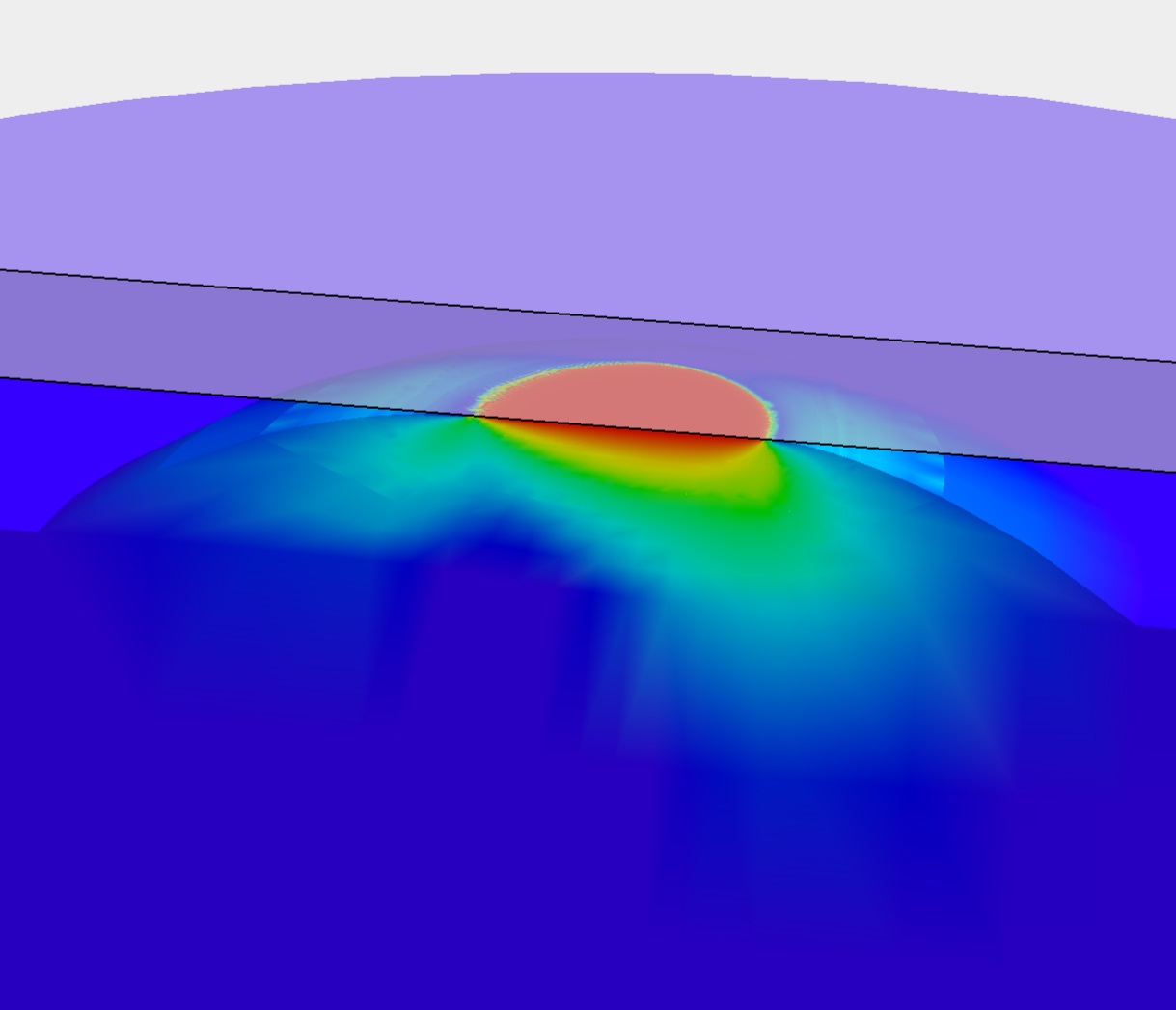}
    \inclps{0.06\textwidth}{!}{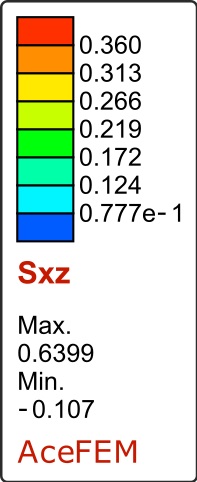}
    &
    \inclps{0.25\textwidth}{!}{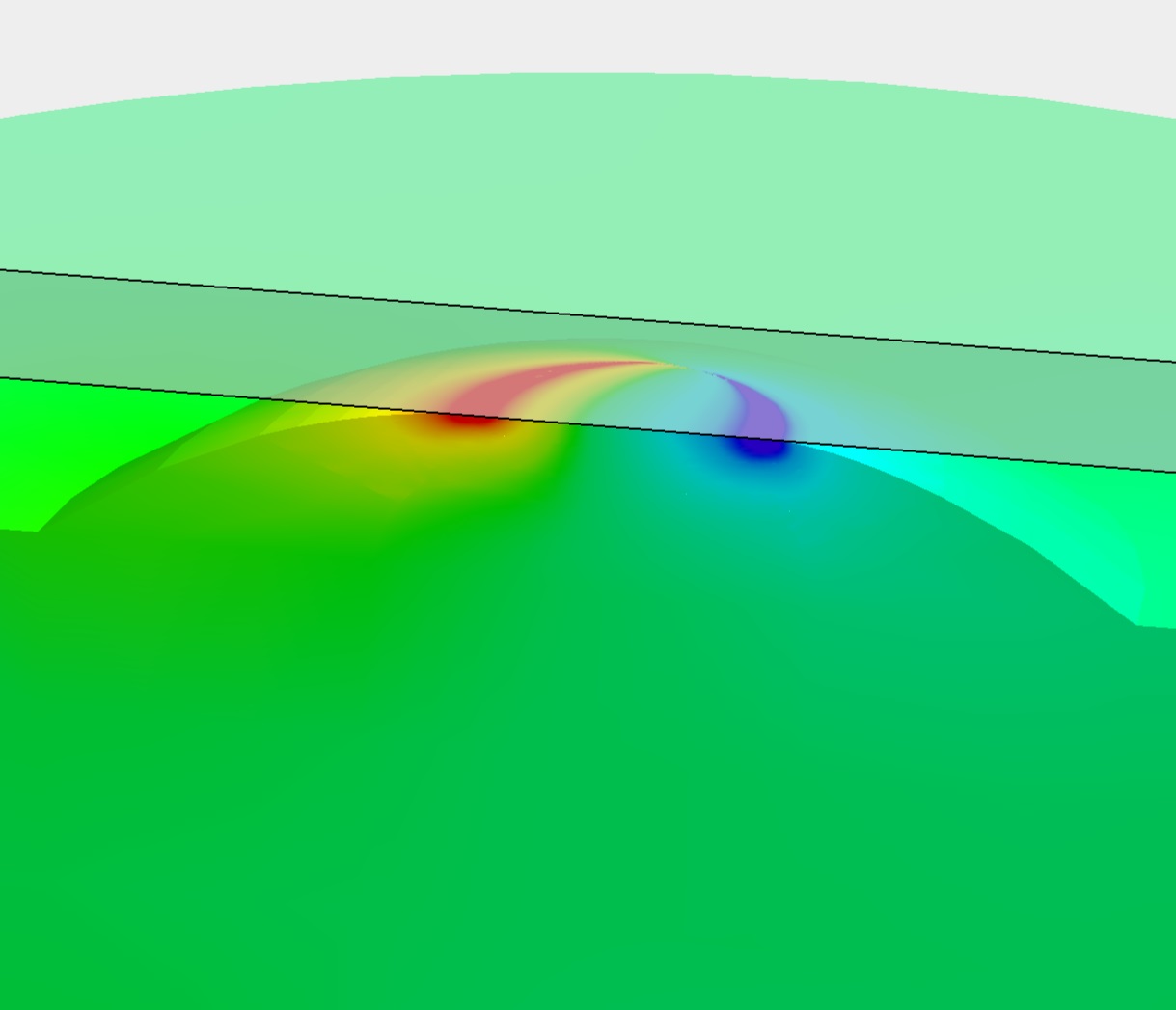}
    \inclps{0.06\textwidth}{!}{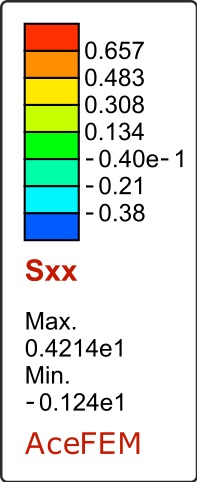} \\[1ex]
    \hspace*{-2em}{\footnotesize (a) $\sigma_{zz}$} & 
    \hspace*{-2em}{\footnotesize (b) $\sigma_{xz}$} & 
    \hspace*{-2em}{\footnotesize (c) $\sigma_{xx}$}
    \end{tabular}
    }
    \caption{Sample finite-element simulations results: components of the Cauchy stress tensor in the deformed configuration at full sliding for the neo-Hookean model. $P=2.12$\,N. All stresses indicated in the boxes are in MPa. {The leading edge is on the left.}}
    \label{fig:3d fem meshes zoomed}
\end{figure}

Sample results of finite-element computations are provided in Fig.~\ref{fig:3d fem meshes zoomed}. The maps of three selected components of the Cauchy stress tensor are shown in the deformed configuration corresponding to full sliding at $P=2.12$\,N. The $\sigma_{zz}$ component is shown in Fig.~\ref{fig:3d fem meshes zoomed}(a). Its value at the contact surface corresponds to the normal contact traction, and the corresponding distribution has a typical Hertz-like appearance. The shear stress $\sigma_{xz}$ is shown in Fig.~\ref{fig:3d fem meshes zoomed}(b) with a constant value at the contact surface, in agreement with the Tresca friction law. Finally, the $\sigma_{xx}$ stress shown in Fig.~\ref{fig:3d fem meshes zoomed}(c) {features} zones of tensile and compressive stresses {corresponding to} surface dilation and compression, respectively, as discussed in detail in Section~\ref{sec:mechanisms}. 

{The contributions of the individual mechanisms of area variations (lifting, laying and in-plane deformations), as reported in Fig.~\ref{fig:Area reduction split plot LS model}, have been determined using} {the following procedure. First, the contact nodes are grouped into six sets that are defined according to the contact state and {local surface dilation/compression}. Set $\mathcal{S}_0$ is determined at $d_0$, the remaining sets are determined at each current displacement $d$.
The following sets are defined:
$\mathcal{S}_0$ -- nodes in contact at $d_0$, i.e., after applying the normal load;
{$\mathcal{S}_d$ -- nodes currently in contact;
$\mathcal{S}_{\text{lift}}$ -- lifted nodes, i.e., those in $\mathcal{S}_0$ but not in $\mathcal{S}_d$;
$\mathcal{S}_{\text{lay}}$ -- new nodes in contact, i.e., those in  $\mathcal{S}_d$ but not in $\mathcal{S}_0$;
$\mathcal{S}_{\text{comp}}$ -- nodes in $\mathcal{S}_0$ and in $\mathcal{S}_d$, for which the tributary area decreased, $A^i<A^i_0$;
$\mathcal{S}_{\text{dil}}$ -- nodes in $\mathcal{S}_0$ and in $\mathcal{S}_d$, for which the tributary area increased, $A^i>A^i_0$.}
The total initial and current areas of the nodes belonging to each set are defined as $a_0(\mathcal{S}) = \sum_{i\in\mathcal{S}} A^i_0$ and $a(\mathcal{S}) = \sum_{i\in\mathcal{S}} A^i$, so that, in particular, $A_0=a_0(\mathcal{S}_0)$ and $A=a(\mathcal{S}_d)$. 
Finally, the contributions to the total relative area change{, as shown in Fig.~\ref{fig:Area reduction split plot LS model},} are defined as: $-a_0(\mathcal{S}_{\text{lift}})/A_0$ (lifting); $a(\mathcal{S}_{\text{lay}})/A_0$ (laying); $(a(\mathcal{S}_{\text{dil}})-a_0(\mathcal{S}_{\text{dil}}))/A_0$ (in-plane dilation); $(a(\mathcal{S}_{\text{comp}})-a_0(\mathcal{S}_{\text{comp}}))/A_0$ (in-plane compression); $(A-A_0)/A_0$ (total reduction).}

\subsection{{Validation of the finite-element model}}
\label{app:validation}

{The present contact} problem is highly demanding {computationally.
{Abrupt changes of contact forces have been discussed in \ref{app:fe-model}.
Moreover, the}} frictional shear strength $\sigma$ is of the same order as the elastic shear modulus $\mu$, and the Tresca model implies that the corresponding high tangential tractions are applied step-wise at the contact zone boundary. 
{As a result, the solution exhibits mesh-dependent features that are visible in Figs.~\ref{fig:contact zone different models 0.27}, \ref{fig:evolution of contact regions glass-plate configuration}, and \ref{fig: In-plane area change glass} and are illustrated in more detail in Section~S.1.2. Additional studies have thus been performed to check the validity of the finite-element results.}

{Firstly, we have carried out additional simulations for a twice coarser mesh, and we have checked that the results are essentially identical to those obtained for the reference mesh and thus, in particular, are not affected by the mesh-dependent features mentioned above, see Section~S.1.2.}

\begin{figure}[ht!]
    \centerline{\inclps{0.41\textwidth}{!}{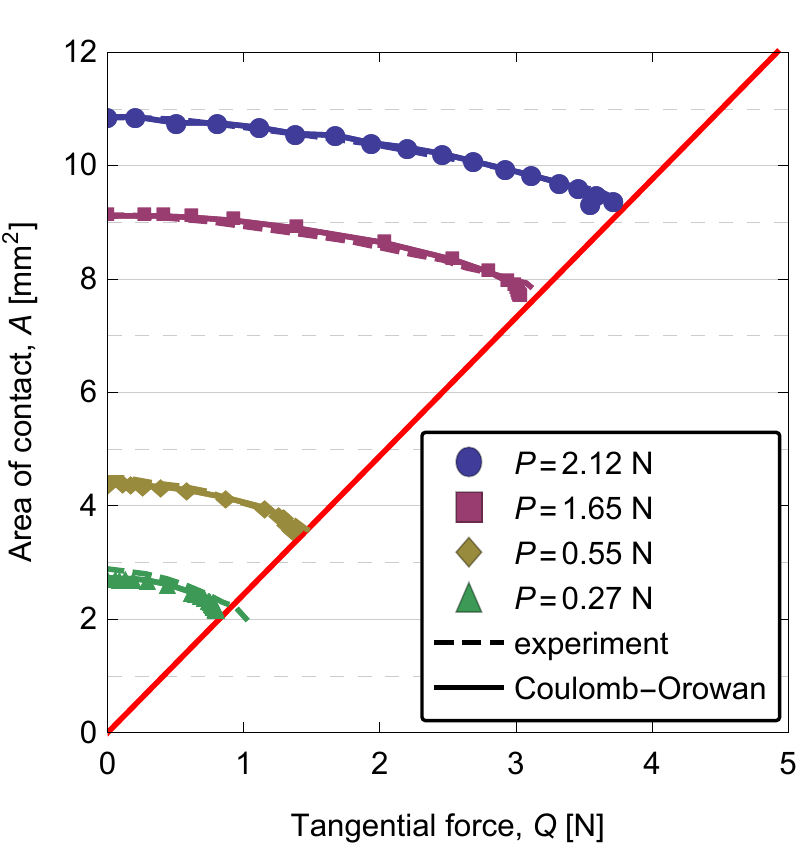}}
    \caption{{Contact area $A$ as a function of the tangential force $Q$, as predicted for the Coulomb--Orowan model (solid lines and symbols).} Dashed lines are the corresponding experimental results from~\citet{sahli_evolution_2018}. {Solid straight red line: $Q=\sigma A$.}}
    \label{fig:ABAQUS}
\end{figure}

{Secondly, we have performed independent computations using the Coulomb--Orowan model (Fig.~\ref{fig:tresca}(c)) with a very high friction coefficient, $f=10$, as an alternative regularization of the Tresca model. The corresponding computations have been carried out using ABAQUS 2020 finite-element package. The details of this alternative model are provided in Section~S.1.3. Sample results are provided in Fig.~\ref{fig:ABAQUS} which shows that the predicted area reduction is in agreement with the experimental data, thus supporting the findings obtained using the main model with Tresca friction.}

\section{Uncertainties on the evaluation of area changes in the experiments}
\label{app:uncertainties}

{We identified 263 tracers strictly larger than 4 pixels in the initial image, on which the tracking procedure has been applied. Out of them, only 31 were useful for the Voronoi analysis (i.e., corresponding either to lifted or laid areas) and 94 for the Delaunay analysis. The average Voronoi area of the useful cells was thus about 0.18\,mm$^2$, i.e., about 1.54\,\% of the initial contact area. In practice, due to the randomness of the tracers' locations, the individual areas varied from 0.2\,\% to 4.0\,\% of the initial contact area. Because the Voronoi cells are the smallest area units in our lifting analysis, the evolution rate of the estimated lifting area suffers from large fluctuations (see drops up to about 3--4\,\% of $A_0$ in the blue curve of Fig.~\ref{fig:Area_Voronoi}). In addition, the final estimate of the lifted area is actually an overestimate because, for the latest lifted cells, the fraction of their area on the left of the tracer has not been really lifted yet. Both effects would be reduced with a larger density of tracers, yielding a smoother curve with a smaller amplitude. The very same discussion holds for our estimate of the laid area.}

{Concerning the estimate of in-plane deformation, {the} region probed by the Delaunay-based analysis is only a fraction of the target region (the part of the initial contact that never lifts nor lays during the experiment). This is apparent for instance on Fig.~\ref{fig:Deformation}(a), with the white strip between the leading part of the contact contour and the Delaunay triangulation. {An} analogous strip is also lost on the trailing side, between the Delaunay triangulation and the lifted region. Those strips are due to the absence of tracers arbitrarily close to the contact contour. Because the largest strains are precisely found on the periphery of the Delaunay triangulation (see Fig.~\ref{fig:Deformation}(d)), the estimate of the area lost by in-plane deformation may be subject to a significant error. The value of 7\,\% of area reduction due to this mechanism must then be taken with caution.}

\section*{Supplementary Information}
\label{app:SI}

\subsection*{S.1: Finite-element model: supplementary information}

\subsubsection*{S.1.1: Finite-element implementation}
\label{sec: finite-element implementation}

For the finite-element discretization of the nearly-incompressible hyperelastic solid at hand, the eight-node hexahedral TSCG12 element \citep{Korelc10} is used. The element employs the enhanced assumed strain (EAS) formulation to circumvent volumetric and shear locking effects and has proven to perform very well in the contact problems considered in this work. 
As a means of verification, selected simulations have been repeated using the F-bar element \citep{SouzaNeto96a}, and the effect of the finite-element formulation has been found negligible.

Consistent with the discretization of the solid, the contact surface is discretized into four-node quadrilateral segments. Nodal quadrature is applied to the contact contribution in the virtual work principle (Eq.~(A.7)) so that, effectively, the contact and friction conditions are evaluated at the individual nodes of the contact surface. In this formulation, the tributary area{, $A^i$,} is determined for each contact {node, of the position $\bm{x}^i$ in the current configuration, according to \citep[cf.][]{LengKorStu11}
\begin{equation}
    A^i = \sum_{k=1}^{K^i} \frac{1}{4} \left\| (\bm{x}^{i,{\rm a}}_k-\bm{x}^i) \times (\bm{x}^{i,{\rm a}}_{k+1}-\bm{x}^i) \right\| ,
\end{equation}
where $\bm{x}^{i,{\rm a}}_k$ denotes the current position of the contact node adjacent to the contact node $\bm{x}^i$ (the adjacent nodes are ordered clockwise), $K^i$ is the number of adjacent nodes ($K^i=4$ for a regular quadrilateral mesh on the contact surface), and we have $\bm{x}^{i,{\rm a}}_{K^i+1}=\bm{x}^{i,{\rm a}}_1$ to simplify the notation. 
The} changes in the contact area can be traced by summing up the tributary areas of the nodes that are in contact at a given instant. Recall that the contact contribution is evaluated in the current configuration, and so is the nodal tributary area. 

The unilateral contact and friction conditions (Eqs.~(A.6) and~(A.8)) are enforced using the augmented Lagrangian method \citep{AlartCurnier91}. In this method, the contact Lagrange multipliers are introduced as global unknowns, and the resulting system of non-linear equations is solved simultaneously with respect to the nodal displacements and Lagrange multipliers using the semi-smooth Newton method. As a result, the contact conditions are enforced in a numerically exact manner.

The beneficial feature of the augmented Lagrangian method is that the contact/separation and stick/slip states are uniquely defined at each global Newton iteration when the contact conditions (Eqs.~(A.6) and~(A.8)) are not yet satisfied. In the time-discretized setting, a special treatment is, however, necessary when switching between the Tresca friction model (Eq.~(A.8)) and its regularization (Eq.~(A.9)). In the direct approach resulting from the implicit time-integration scheme, the \emph{current} contact/separation state determined at each iteration is used to choose whether the Tresca model (in the case of contact) or the regularization scheme (in the case of separation) is employed. However, we have observed that this leads to severe convergence problems. Accordingly, solely with the aim to determine whether to use the Tresca model or its regularization, the contact/separation state is here taken from the \emph{last converged} time step. Furthermore, an adaptive time-stepping procedure is applied, which conveniently alleviates the difficulties caused by the high non-linearity of the problem. 

Computer implementation has been performed using \emph{AceGen}, a code generation system that employs the automatic differentation (AD) technique \citep{Korelc09,KorelcWriggers16}, see also \citet{LengKorStu11} for the details of the automation of the finite-element contact formulations. The computations have been performed in \emph{AceFEM}, a flexible finite-element environment that is interfaced with \emph{AceGen}.

\subsubsection*{S.1.2: Supplementary results of finite-element computations}

{Further comparison of the neo-Hookean model predictions with the experimental results of \citet{sahli_evolution_2018} is provided in Fig.~S.1 which shows the concurrent evolution of the tangential force, $Q$, and the contact area, $A$, as a function of the rigid plate displacement, $d$. The results correspond to those presented in Fig.~3(a) which shows $A$ as a function of $Q$. It can be seen that the dependence of $Q$ on $d$ is captured by the model with a good accuracy, particularly at the initial stage of loading.}

\begin{figure}[htb!]
    \centerline{
    \begin{tabular}{cc}
    \hspace*{-1em}
    \inclps{0.48\textwidth}{!}{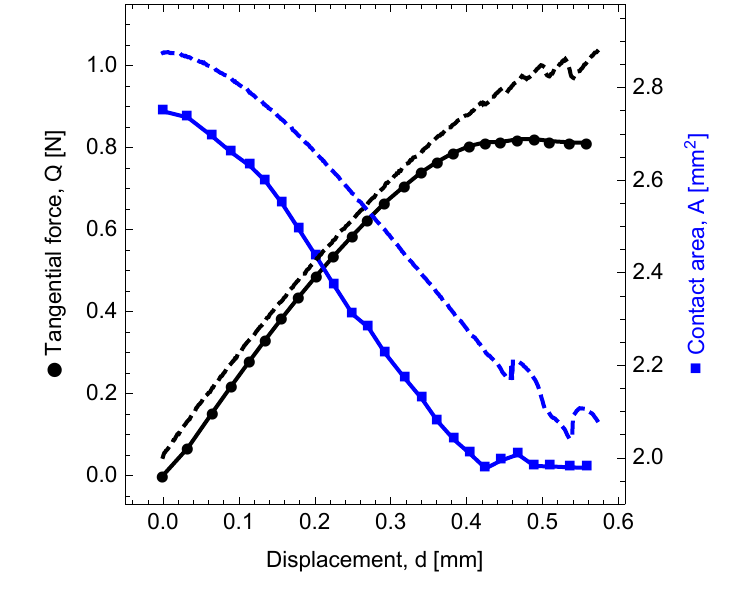} \hspace*{-1em}
    &
    \hspace*{-1em}
    \inclps{0.48\textwidth}{!}{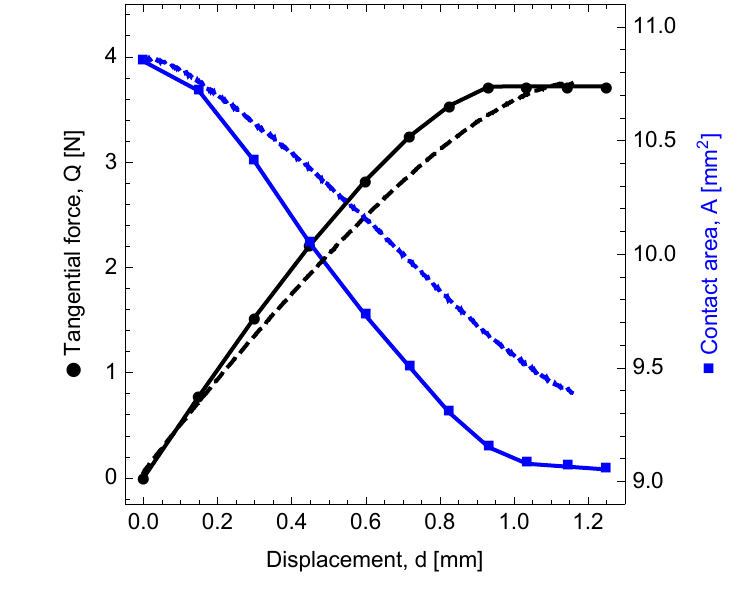}
    \hspace*{-1em}
    \\[-1ex]
    {\footnotesize (a)} &
    {\footnotesize (b)}
    \end{tabular}
    }
		\captionsetup{labelformat=empty}
    \caption{Figure S.1: {Concurrent evolution of the tangential force, $Q$ (dots), and the contact area, $A$ (squares), as a function of the rigid plate displacement, $d$, for the neo-Hookean model, and for: (a) $P=0.27$\,N, (b) $P=2.12$\,N. Results of finite-element computations (solid lines) are compared to the experiment of \citet{sahli_evolution_2018} (dashed lines).}}
    \label{fig:evolution}
\end{figure}

{As discussed in the paper, the Tresca friction model with a high contact shear strength $\sigma$ may lead to a significant distortion of the finite-element mesh. As a result,} the boundary of the contact zone may exhibit mesh-dependent features, {like those} illustrated in Fig.~S.2(a). Mesh-dependent features in the form of secondary contact regions are also observed, see Fig.~S.2(b), particularly at lower normal load $P$ and higher tangential load $Q$ (cf.\ Fig.~6). Since those separated contact spots have a small total area compared to that of the main contact region, they have been included in all evaluations of the total contact area, $A$. In contrast, when we evaluated the contact sizes along and orthogonal to shear, $\ell_{\parallel}$ and $\ell_{\perp}$ respectively, those separated spots have been excluded because they would have induced a significant noise and error, especially on $\ell_{\parallel}$ (see e.g. Fig.~S.2(b)).

\begin{figure}[htb!]
    \centerline{
    \hspace*{0.\textwidth}
    \begin{tabular}{cc}
    \raisebox{3ex}{\inclps{0.26\textwidth}{!}{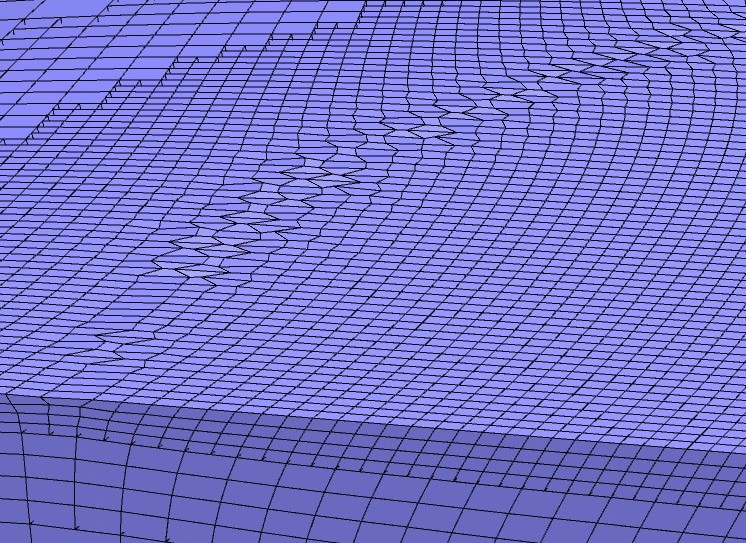}}
    \inclps{0.22\textwidth}{!}{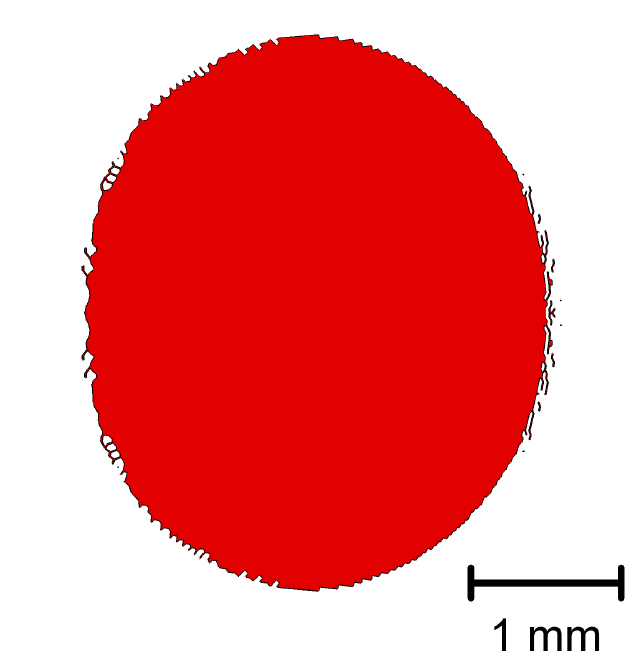} &
    \inclps{0.22\textwidth}{!}{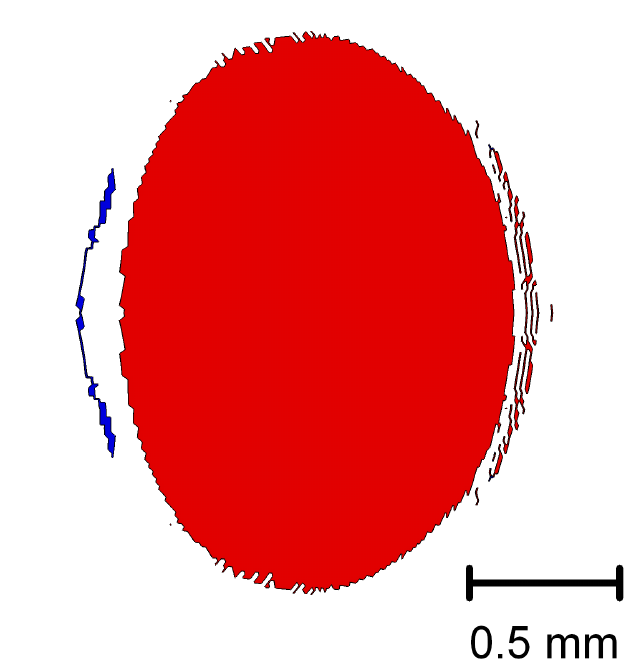}\hspace*{-1em}
    \inclps{0.22\textwidth}{!}{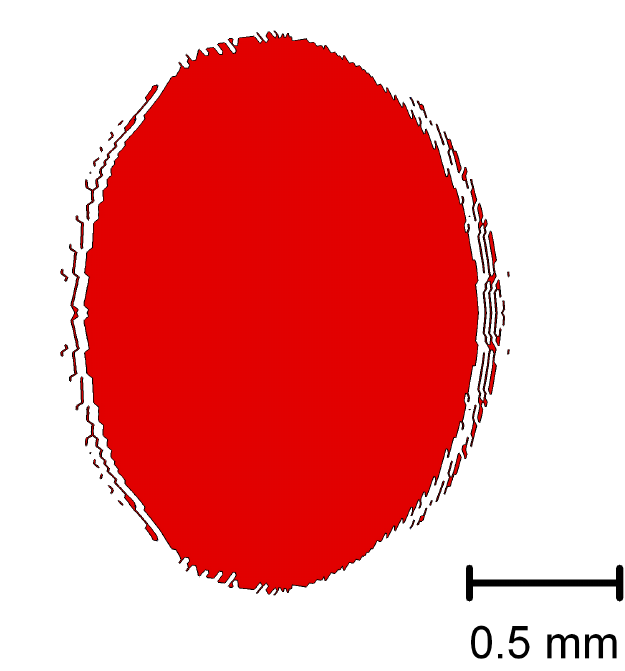} \\[1ex]
    {\footnotesize (a)}~~~ &
    {\footnotesize (b)}
    \end{tabular}
    }
		\captionsetup{labelformat=empty}
    \caption{Figure S.2: Illustration of mesh-dependent features associated with the Tresca model and high contact shear strength $\sigma$. (a) Left: expanded view of Fig.~2(d) showing mesh distortion at the leading edge at $P=2.12$\,N ; right: corresponding contact area with a wiggled periphery. (b) Contact area at two instants during gross sliding at $P=0.27$\,N showing small contact spots developing and evolving outside the main contact area.}
    \label{fig:artifacts}
\end{figure}

With the aim to validate our results, we have carried out additional simulations using a finite-element mesh twice coarser than that used in the main computations. We have checked that the results are essentially identical to those obtained for the reference mesh and thus are not affected by the mesh-dependent features discussed {above}. This, in particular, concerns the evolution of the contact area as a function of the tangential force, as shown in Fig.~S.3. Due to the high computational cost and significant convergence challenges, we were not able to solve the problem for a twice finer mesh. 

\begin{figure}[ht!]
    \centerline{
	\inclps{0.4\textwidth}{!}{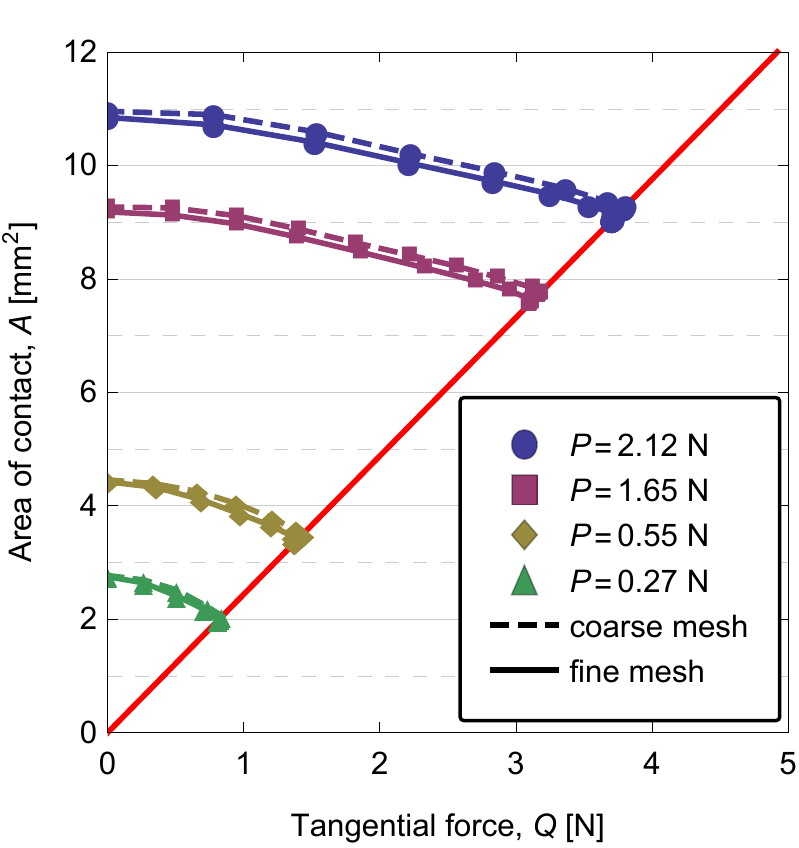}
    }
		\captionsetup{labelformat=empty}
    \caption{Figure S.3: Contact area $A$ as a function of the tangential force $Q$: effect of the mesh density. 
    Red solid straight line: $Q=\sigma A$.}
    \label{fig:check}
\end{figure}

\subsubsection*{S.1.3: Alternative model using Coulomb-Orowan friction}
\label{app:reg-coulomb-orowan}

To test the robustness of the results obtained using the time-regularized Tresca friction model (Eq.~(A.9)) additional computations have been carried out using the Coulomb--Orowan model (Fig.~A.1(c)) as an alternative regularization approach. The friction coefficient in the Coulomb--Orowan model has been adopted equal to $f=10$, which in practice was the highest achievable value leading to convergence. All geometrical and material parameters and boundary conditions are otherwise the same as in the main (Tresca-based) model.

The unilateral contact and friction conditions were enforced using the penalty method. In particular, the tangential tractions could reach the contact shear strength, $\sigma$, only when a slip distance of 0.3\,\% of the contact size was reached. Such an elastic slip allowed to smoothly reach the final shear state and to converge faster. Using this approach, the evolution of the contact area versus the tangential load can be consistently extracted, where the intermediate states correspond to a progressive transition of the contact zone from the sticking to sliding state.

The corresponding computations have been carried out using ABAQUS 2020 finite-element package. The hyperelastic (neo-Hookean) solid was meshed with 8-node linear brick hybrid elements with reduced integration (C3D8RH). In order to alleviate the issues related to the almost incompressible nature of the material and volumetric locking problems, the hybrid formulation was used~\citep{nguyen_surface_2011}. The mesh was refined down to 30\,$\mu$m in the vicinity of the contact zone for all normal loads, leading to a total of nearly 145,000 hexahedral elements and about 580,000 degrees of freedom, see Fig.~S.4(a).

\begin{figure}[ht!]
    \centerline{
    {\footnotesize (a)}~~~\inclps{0.43\textwidth}{!}{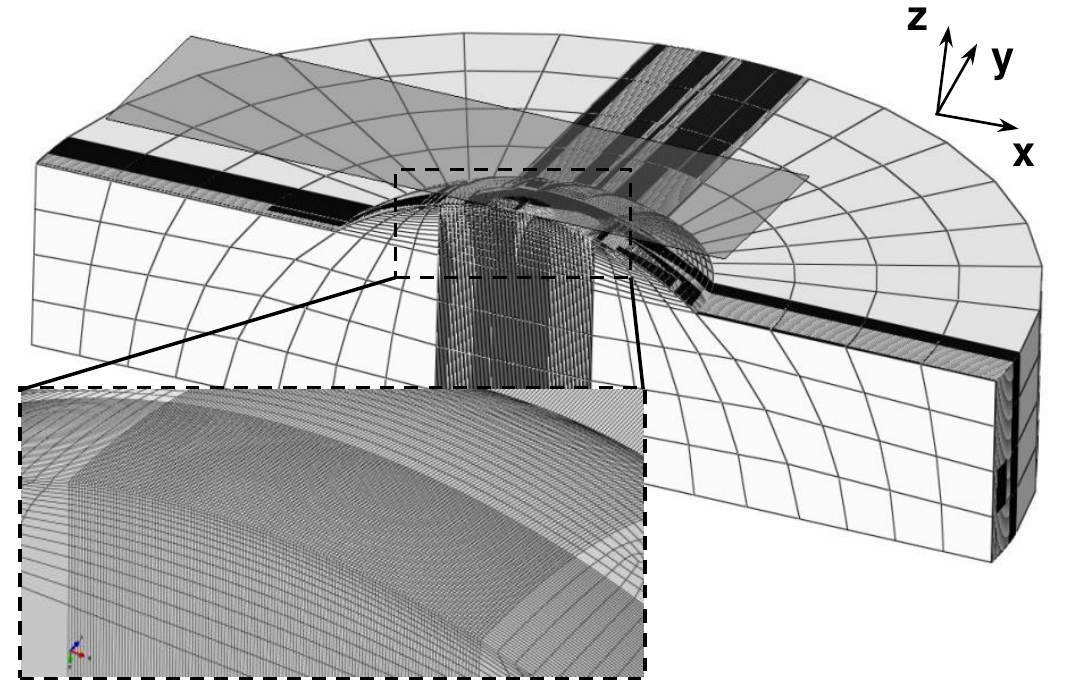}
    \hspace*{2em}
    {\footnotesize (b)}\hspace*{0em}\inclps{0.30\textwidth}{!}{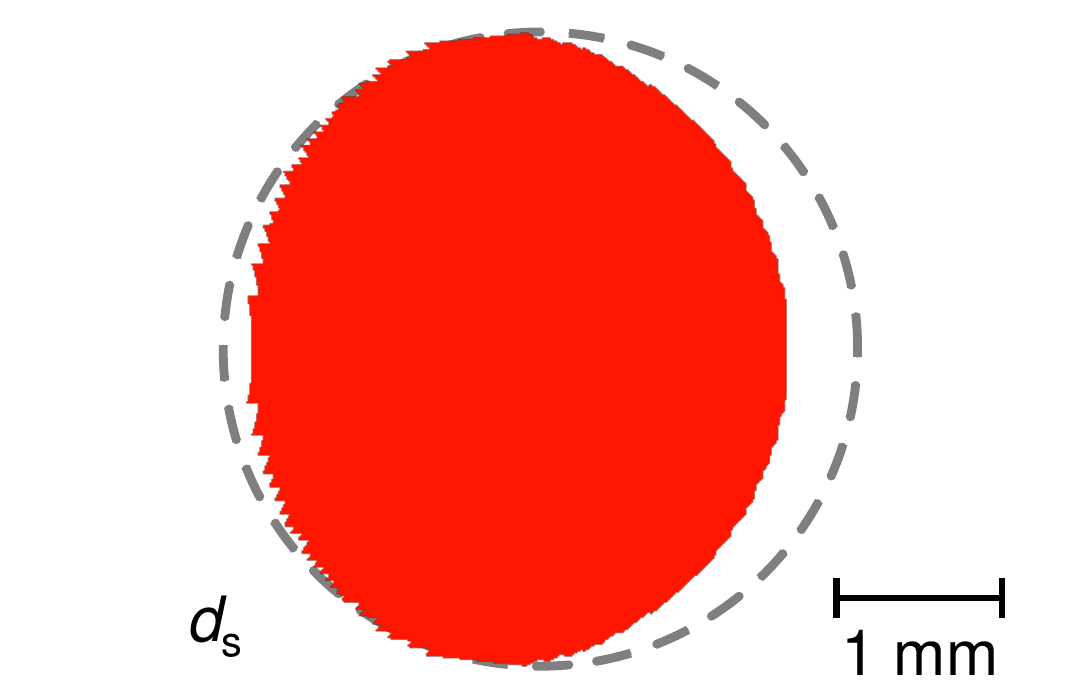}
    }
		\captionsetup{labelformat=empty}
		\caption{Figure S.4: Sample results obtained for the Coulomb--Orowan model. (a) Views of the mesh. (b) Contact morphology at $d_{\rm s}$ (red) compared to the initial contact (dashed circle), for $P=2.12$\,N.}
    \label{fig:check_ABQ}
\end{figure}

The large-displacement formulation was employed to account for the various non-linearities and solved using an implicit integration scheme based on the Newton--Raphson method. To overcome convergence problems, the simulation was perfomed in two steps. In the first step, the normal load $P$ was applied in frictionless conditions. In the second step, the sliding velocity and the friction conditions were applied. 

From a qualitative point of view, the predicted shape of the contact morphology, shown in Figure~S.4(b) for the full sliding configuration, is found consistent with that of the Tresca model (compare to Fig.~4(b)). An anisotropic change is also observed with a stronger reduction at the trailing edge of the contact. Figure~A.3 summarizes the quantitative evolution of the contact area for the same four loading cases considered in the Tresca model. The predicted area reduction is in good agreement with the experimental data and thus also with that obtained using the Tresca model (Fig.~5(a)). Such a good correlation shows that the main model results are robust against the details of the regularization method adopted.

The main limitation of the Colomb--Orowan model is that the contact shear strength $\sigma$ cannot be reached at the very periphery of the contact, where the local pressure remains smaller than $\sigma / f$. In this region, the limiting shear stress is set by the friction coefficient $f$ rather than by the contact shear strength $\sigma$. As a consequence, the maximum tangential force accessible in the simulations remains smaller than the product of $\sigma$ with the contact area, and thus the simulated curves do not reach perfectly the desired maximum tangential force, denoted as the red line in Fig.~A.3.

\subsection*{S.2: Detailed experimental methods}

\subsubsection*{S.2.1: Opto-mechanical test}\label{app:meca}

We used a laboratory-built experimental setup in which the elastomer sample is fixed to an optical table, while the glass substrate is attached at the extremity of a horizontal cantilever beam composed of two identical arms. The cantilever has a stiffness $177\pm1$\,N/m in the direction normal to the contact interface and is essentially rigid in the direction parallel to the interface. To shear the elastomer/glass interface, the other extremity of the cantilever is pulled horizontally by a linear motor (Newport LTA-HL) at a constant velocity $V$. The tangential force $Q$ is measured via a piezoelectric sensor (Kistler 9217A) placed between the motor and the cantilever. The normal force $P$ is constant and applied by adding a dead weight to the configuration in which the elastomer sphere just snapped into contact with the glass due to adhesion. The latter configuration is reached by slowly translating vertically the whole cantilever. The value of $P$ is measured via four load cells (Futek) localized between the table and the elastomer sample. The tangential and normal forces are digitized and recorded at a rate of 2\,kHz with, respectively, 5\,mN and 0.2\,mN resolutions.

The contact interface is illuminated from above with a white LED panel and the contact area is monitored optically in reflection mode, at 100 frames per second, using a CCD camera (Flare 2M360 MCL, 8 bits, 2048$\times$1088 square pixels) attached to a variable-zoom objective lens (Navitar 7000 type C). In the conditions of the experiments, each pixel corresponds to a square of lateral size of 7.70\,$\mu$m in the contact plane. True contact corresponds to dark pixels (see e.g. Fig.~12(a--d)), because the light rays transmitted through the PDMS/glass interface are absorbed by a black layer inserted at the bottom of the sample. The lighter gray background corresponds to out-of-contact regions where a fraction of the light intensity is reflected back to the camera, at the interface between the bottom face of the glass substrate and the air.

A homemade LabVIEW (National Instruments) routine was used to record the acquisitions of the forces, $P$ and $Q$, and a homemade MATLAB (The MathWorks, Inc.) software was used to synchronize the data with the images.

\subsubsection*{S.2.2: Elastomer and glass samples}\label{app:sample}

The particle-filled cross-linked PolyDiMethylSiloxane (PDMS, Sylgard 184, Dow Corning) sample with a spherical cap was prepared as follows. The elastomer base and curing agent are respectively mixed in a 10:1 weight ratio and degassed in a vacuum chamber to remove air bubbles introduced during mixing. Cross-linking is performed in two steps. First, a drop of PDMS mixture is poured into a plano-concave glass lens with a radius of curvature of 9.42\,mm (Edmund Optics), spin-coated (SPIN 150, Spincoating) for 30\,s at 2500\,rpm and cured in an oven (Prolabo, Astel S.A.) at $80^\circ$C for 20\,min, to obtain a thin cross-linked PDMS layer, approximately 16\,$\mu$m thick. The exposed surface of this layer is then sprinkled with silver particles, obtained from drying an Electrodag 1415M (Agar Scientific Ltd) diluted in acetone. In a second step, the lens with the particle-seeded PDMS layer is placed at the bottom of a cylindrical cavity (30\,mm in diameter), which is then filled with the PDMS mixture, covered with a glass plate (40\,mm$\times$35\,mm$\times$5\,mm) and cured at room temperature for 48\,h. Once demolded, the sample has a 6\,mm-high cylinder-like shape, the top of which features a spherical cap of a radius of curvature 9.42\,mm, with a summit 2\,mm higher than the top surface of the cylinder (see inset of Fig.~10).

{The Young's modulus of the elastomer has been measured to be $E=1.5 \pm 0.1$\,MPa from a fit of $A_0$, the area of contact under pure normal load $P$, as a function of $P$ (in the range 0.05--1.85\,N), using Johnson-Kendall-Roberts' model~\citep{johnson_surface_1971}.}

A 5\,mm-thick smooth bare-glass plate (Mirit Glass) was used as a slider. Before experiments, the glass surface was first gently pre-cleaned mechanically with distilled water using a lens cleaning tissue. It was then left in three different ultrasonic baths, each for 15\,min, in the following order: soapy water (Decon90), ethanol and distilled water. Between the baths, the plate was rinsed in distilled water. It was finally dried in an oven at $90^\circ$C for 15\,min.

\subsubsection*{S.2.3: Image analysis: segmentation and particle tracking}
\label{app:image}

The first step in the image analysis is the segmentation of the raw images into in- and out-of-contact pixels. For this, we first crop the images around the contact, thus excluding the marker on the glass substrate (see top right black spot in Figs.~12(a-d)). Then, we transform the raw gray-scale cropped images into binary images by thresholding. The threshold value is kept constant for all images and is selected by applying Otsu's method~\citep{otsu_threshold_1979} to the initial image. The segmented images can be directly used to measure the contact area, $A$, through selection of the largest dark object in the image and filling up of its inner holes, due to the presence of the particles. Typical contours of the contact are shown in Figs.~12(e-h).

The second step consists in the following denoising procedure applied to all images: (i) all objects whose area is a single pixel are removed, then (ii) the closest remaining objects are connected by successively performing one dilation and one erosion operation with a 3$\times$3 square structuring element.

During the third step, we identify, in the initial image, all the objects whose area is strictly larger than 4 pixels. Those objects from the initial image are now considered as the tracers of the contact evolution for upcoming analysis. Note that for each tracer, we store the $x$- and $y$-widths of the smallest rectangle enclosing the tracer ($W_x$ and $W_y$).
  
The fourth step is to find the trajectory of each tracer by searching  its successive positions in all  images. For each couple of two consecutive images, the following tracking procedure is applied for each tracer:
\begin{itemize}
        \item extract two sub-images (first and second) from both consecutive images: their common position corresponds to the location of the tracer in the first image and their common widths correspond to $W_x$ and $W_y$ increased by 5 pixels on all four sides, to conservatively account for the dilation operation that will be performed in the next step. 
        \item perform a  dilation  operation with a 3$\times$3 square structuring element on the first sub-image. This dilation has an amplitude much larger than the distance ($\Delta d_{\rm max}$) traveled by the slider between two images in steady sliding regime, $\Delta d_{\rm max}=1\,\mu{\rm m} \approx 0.14$ pixel.
        \item multiply pixel by pixel the two resulting sub-images, i.e., the dilated first one with the unmodified second one.
        \item find all objects (connected components) in the multiplication image. Store the position of the object as the coordinates $(T_x,T_y)$ of its center of mass in the full image. In the case of multiple objects, only the closest object to the center of the sub-image is considered. If there is no object, then the trajectory ends.
\end{itemize}

The fifth step consists in removing all the trajectories that contain a non-realistic change, greater than 3 pixels, in the position of a tracer between two successive images.
 
In the last step, we first measure the glass plate displacement by tracking the position of the centroid of the tracer drawn on the glass for each image. We then translate the segmented and denoised counterparts of the raw images (Fig.~12(a-d)) in such a way that this tracer remains at a constant position.  In other words, we place our images in the frame of the glass substrate (Fig.~12(e-h)), which actually moves from left to right at velocity $V$ in the laboratory frame. Finally, we subtract the glass plate displacement to all our trajectories.

The outcome of the tracking analysis is thus a dataset in which the position $(T_x(t),T_y(t))$ of each tracer is given in the glass frame, for each image (i.e., for each time step).

\section*{References}



\end{document}